\RequirePackage{fix-cm}
\RequirePackage{rotating}
\documentclass{svjour3}                     % onecolumn (standard format)
\makeatletter
 \def\@textbottom{\vskip \z@ \@plus 1pt}
 \let\@texttop\relax
\makeatother
\smartqed  % flush right qed marks, e.g. at end of proof
\usepackage{hyperref}
\usepackage{graphicx}
\usepackage{verbatim}
\usepackage{subfig}
\usepackage[table,xcdraw]{xcolor}
\usepackage{multirow}
\usepackage{amsmath,amssymb}                % amsmath and amssymb packages, useful for mathematical formulas and symbols
\usepackage[aboveskip=1pt,labelfont=bf,labelsep=space,justification=raggedright,singlelinecheck=off]{caption}           % Bold the 'Figure #' in the caption and separate it from the title/caption with a period; Captions will be left justified
\usepackage[toc,page]{appendix}
\usepackage{booktabs}
\usepackage{rotating,tabularx}
\usepackage{footnote}
\makesavenoteenv{tabular}
\makesavenoteenv{table}
\usepackage[utf8]{inputenc}
\definecolor{cecccc}{HTML}{cecccc}
\definecolor{fffdfd}{HTML}{fffdfd}
\definecolor{ff0000}{HTML}{ff0000}
\definecolor{ffb1b1}{HTML}{ffb1b1}
\definecolor{ff6262}{HTML}{ff6262}
\definecolor{ff3a3a}{HTML}{ff3a3a}
\definecolor{ffffff}{HTML}{ffffff}
\definecolor{ff0404}{HTML}{ff0404}
\definecolor{ffc3c3}{HTML}{ffc3c3}
\definecolor{ff7474}{HTML}{ff7474}
\definecolor{ff3e3e}{HTML}{ff3e3e}
\definecolor{fff7f7}{HTML}{fff7f7}
\definecolor{ff0909}{HTML}{ff0909}
\definecolor{ffa5a5}{HTML}{ffa5a5}
\definecolor{ff4646}{HTML}{ff4646}
\definecolor{ff3030}{HTML}{ff3030}
\definecolor{ffe6e6}{HTML}{ffe6e6}
\definecolor{ff3434}{HTML}{ff3434}
\definecolor{ffd9d9}{HTML}{ffd9d9}
\definecolor{ff1818}{HTML}{ff1818}
\definecolor{ffe1e1}{HTML}{ffe1e1}
\definecolor{ff3b3b}{HTML}{ff3b3b}
\definecolor{ff4141}{HTML}{ff4141}
\definecolor{ffffff}{HTML}{ffffff}
\definecolor{ffcdcd}{HTML}{ffcdcd}
\definecolor{ff0505}{HTML}{ff0505}
\definecolor{ffb9b9}{HTML}{ffb9b9}
\definecolor{ff6d6d}{HTML}{ff6d6d}
\definecolor{ff3f3f}{HTML}{ff3f3f}
\definecolor{ffebeb}{HTML}{ffebeb}
\definecolor{ff2e2e}{HTML}{ff2e2e}
\definecolor{ff9f9f}{HTML}{ff9f9f}
\definecolor{ffb5b5}{HTML}{ffb5b5}
\definecolor{ff9090}{HTML}{ff9090}
\definecolor{fffcfc}{HTML}{fffcfc}
\definecolor{ff6969}{HTML}{ff6969}
\definecolor{ff2828}{HTML}{ff2828}
\definecolor{fff9f9}{HTML}{fff9f9}
\definecolor{ff6f6f}{HTML}{ff6f6f}
\definecolor{ffaaaa}{HTML}{ffaaaa}
\definecolor{ff7a7a}{HTML}{ff7a7a}
\definecolor{ff5b5b}{HTML}{ff5b5b}
\definecolor{ff2323}{HTML}{ff2323}
\definecolor{ff9d9d}{HTML}{ff9d9d}
\definecolor{ff6868}{HTML}{ff6868}
\definecolor{ff4d4d}{HTML}{ff4d4d}
\definecolor{fffbfb}{HTML}{fffbfb}
\definecolor{ff7979}{HTML}{ff7979}
\definecolor{ff2121}{HTML}{ff2121}
\definecolor{ffcaca}{HTML}{ffcaca}
\definecolor{ff6060}{HTML}{ff6060}
\definecolor{ff3232}{HTML}{ff3232}
\definecolor{ffdfdf}{HTML}{ffdfdf}
\definecolor{ffa7a7}{HTML}{ffa7a7}
\definecolor{ff3939}{HTML}{ff3939}
\definecolor{ff1919}{HTML}{ff1919}
\definecolor{ffe4e4}{HTML}{ffe4e4}
\definecolor{ff6e6e}{HTML}{ff6e6e}
\definecolor{ff2c2c}{HTML}{ff2c2c}
\definecolor{fff8f8}{HTML}{fff8f8}
\definecolor{ff1616}{HTML}{ff1616}
\definecolor{ffa9a9}{HTML}{ffa9a9}
\definecolor{ff2a2a}{HTML}{ff2a2a}
\definecolor{ff0606}{HTML}{ff0606}
\definecolor{ffd1d1}{HTML}{ffd1d1}
\definecolor{ff7a7a}{HTML}{ff7a7a}
\definecolor{fffefe}{HTML}{fffefe}
\definecolor{ffafaf}{HTML}{ffafaf}
\definecolor{ff6666}{HTML}{ff6666}
\definecolor{ffd8d8}{HTML}{ffd8d8}
\definecolor{ffdcdc}{HTML}{ffdcdc}
\definecolor{ff7272}{HTML}{ff7272}
\definecolor{ffbdbd}{HTML}{ffbdbd}
\definecolor{ffdddd}{HTML}{ffdddd}
\definecolor{fff2f2}{HTML}{fff2f2}
\definecolor{ff9e9e}{HTML}{ff9e9e}
\definecolor{ff3535}{HTML}{ff3535}
\definecolor{ff8686}{HTML}{ff8686}
\definecolor{fff6f6}{HTML}{fff6f6}
\definecolor{ff1212}{HTML}{ff1212}
\definecolor{ffcbcb}{HTML}{ffcbcb}
\definecolor{ff6c6c}{HTML}{ff6c6c}
\definecolor{ffc2c2}{HTML}{ffc2c2}
\definecolor{ff3636}{HTML}{ff3636}
\definecolor{ffa3a3}{HTML}{ffa3a3}
\definecolor{ff0202}{HTML}{ff0202}
\definecolor{ffcfcf}{HTML}{ffcfcf}
\definecolor{ffe7e7}{HTML}{ffe7e7}
\definecolor{ff2424}{HTML}{ff2424}
\definecolor{ffb8b8}{HTML}{ffb8b8}
\definecolor{ffc1c1}{HTML}{ffc1c1}
\definecolor{ffb3b3}{HTML}{ffb3b3}
\definecolor{ff6464}{HTML}{ff6464}
\definecolor{ffd3d3}{HTML}{ffd3d3}
\definecolor{fffafa}{HTML}{fffafa}
\definecolor{ff6363}{HTML}{ff6363}
\definecolor{ffc8c8}{HTML}{ffc8c8}
\definecolor{ff6a6a}{HTML}{ff6a6a}
\definecolor{ffdada}{HTML}{ffdada}
\definecolor{ff5c5c}{HTML}{ff5c5c}
\definecolor{ff8c8c}{HTML}{ff8c8c}
\definecolor{ff8181}{HTML}{ff8181}
\definecolor{ff7777}{HTML}{ff7777}
\definecolor{fff1f1}{HTML}{fff1f1}
\definecolor{ff5d5d}{HTML}{ff5d5d}
\definecolor{ffabab}{HTML}{ffabab}
\definecolor{ff9595}{HTML}{ff9595}
\definecolor{ff5a5a}{HTML}{ff5a5a}
\definecolor{ff8585}{HTML}{ff8585}
\definecolor{ffe9e9}{HTML}{ffe9e9}
\definecolor{ff1717}{HTML}{ff1717}
\definecolor{ff2727}{HTML}{ff2727}
\definecolor{ff0a0a}{HTML}{ff0a0a}
\definecolor{ffecec}{HTML}{ffecec}
\definecolor{ff2222}{HTML}{ff2222}
\definecolor{ff0101}{HTML}{ff0101}
\definecolor{ffe0e0}{HTML}{ffe0e0}
\definecolor{ff8080}{HTML}{ff8080}
\definecolor{ff4b4b}{HTML}{ff4b4b}
\definecolor{ffacac}{HTML}{ffacac}
\definecolor{ffbcbc}{HTML}{ffbcbc}
\definecolor{ffb4b4}{HTML}{ffb4b4}
\definecolor{ffc4c4}{HTML}{ffc4c4}
\definecolor{ff9c9c}{HTML}{ff9c9c}
\definecolor{ffa4a4}{HTML}{ffa4a4}
\definecolor{ff0808}{HTML}{ff0808}
\definecolor{ffc0c0}{HTML}{ffc0c0}
\definecolor{ff8484}{HTML}{ff8484}
\definecolor{ff3c3c}{HTML}{ff3c3c}
\definecolor{ff4343}{HTML}{ff4343}
\definecolor{ff9696}{HTML}{ff9696}
\definecolor{ff3333}{HTML}{ff3333}
\definecolor{ff5656}{HTML}{ff5656}
\definecolor{ff5e5e}{HTML}{ff5e5e}
\definecolor{ff4848}{HTML}{ff4848}
\definecolor{ffd4d4}{HTML}{ffd4d4}
\definecolor{ff9a9a}{HTML}{ff9a9a}
\definecolor{ffdede}{HTML}{ffdede}
\definecolor{ffbbbb}{HTML}{ffbbbb}
\definecolor{fff4f4}{HTML}{fff4f4}
\definecolor{ffefef}{HTML}{ffefef}
\definecolor{ff7e7e}{HTML}{ff7e7e}
\definecolor{ff5151}{HTML}{ff5151}
\definecolor{ff1010}{HTML}{ff1010}
\definecolor{ffcccc}{HTML}{ffcccc}
\definecolor{ffbaba}{HTML}{ffbaba}
\definecolor{ffc6c6}{HTML}{ffc6c6}
\definecolor{ffaeae}{HTML}{ffaeae}
\definecolor{ff5454}{HTML}{ff5454}
\definecolor{ffc5c5}{HTML}{ffc5c5}
\definecolor{ff4f4f}{HTML}{ff4f4f}
\definecolor{ff7373}{HTML}{ff7373}
\definecolor{ff3737}{HTML}{ff3737}
\definecolor{ffeded}{HTML}{ffeded}
\definecolor{ff7171}{HTML}{ff7171}
\definecolor{ffa1a1}{HTML}{ff9191}
\definecolor{ffe2e2}{HTML}{ffe2e2}
\definecolor{fff0f0}{HTML}{fff0f0}
\definecolor{ffadad}{HTML}{ffadad}
\definecolor{ff3838}{HTML}{ff3838}
\definecolor{fff3f3}{HTML}{fff3f3}
\definecolor{ffc9c9}{HTML}{ffc9c9}
\definecolor{ff0b0b}{HTML}{ff0b0b}
\definecolor{ffd5d5}{HTML}{ffd5d5}
\definecolor{ffe5e5}{HTML}{ffe5e5}
\definecolor{ff6b6b}{HTML}{ff6b6b}
\definecolor{ff9494}{HTML}{ff9494}
\definecolor{ff5252}{HTML}{ff5252}
\definecolor{ff5959}{HTML}{ff5959}
\definecolor{ffb6b6}{HTML}{ffb6b6}
\definecolor{ffbebe}{HTML}{ffbebe}
\definecolor{ff1414}{HTML}{ff1414}
\definecolor{ffb2b2}{HTML}{ffb2b2}
\definecolor{ff9292}{HTML}{ff9292}
\definecolor{ff5858}{HTML}{ff5858}
\definecolor{ff8a8a}{HTML}{ff8a8a}
\definecolor{ff5050}{HTML}{ff5050}
\definecolor{ffcece}{HTML}{ffcece}
\definecolor{ff9999}{HTML}{ff9999}
\definecolor{ffd0d0}{HTML}{ffd0d0}
\definecolor{ff1d1d}{HTML}{ff1d1d}
\definecolor{ffa6a6}{HTML}{ffa6a6}
\definecolor{ffd2d2}{HTML}{ffd2d2}
\definecolor{ff7f7f}{HTML}{ff7f7f}
\definecolor{ffdbdb}{HTML}{ffdbdb}
\definecolor{ff8383}{HTML}{ff8383}
\definecolor{ff5757}{HTML}{ff5757}
\definecolor{ff7c7c}{HTML}{ff7c7c}
\definecolor{ff8989}{HTML}{ff8989}
\definecolor{ff7878}{HTML}{ff7878}
\definecolor{ff8282}{HTML}{ff8282}
\definecolor{ff0c0c}{HTML}{ff0c0c}
\definecolor{ff4949}{HTML}{ff4949}
\definecolor{ff9898}{HTML}{ff9898}
\definecolor{ffc7c7}{HTML}{ffc7c7}
\definecolor{ff3d3d}{HTML}{ff3d3d}
\definecolor{ff4747}{HTML}{ff4747}
\definecolor{ff3131}{HTML}{ff3131}
\definecolor{ff7575}{HTML}{ff7575}
\definecolor{ff6565}{HTML}{ff6565}
\definecolor{ff0d0d}{HTML}{ff0d0d}
\definecolor{ff1313}{HTML}{ff1313}
\definecolor{ff1f1f}{HTML}{ff1f1f}
\definecolor{ff1b1b}{HTML}{ff1b1b}
\definecolor{ff9191}{HTML}{ff9191}
\definecolor{ff5f5f}{HTML}{ff5f5f}
\definecolor{ff2626}{HTML}{ff2626}
\definecolor{ff0707}{HTML}{ff0707}
\definecolor{ff4c4c}{HTML}{ff4c4c}
\definecolor{ffd7d7}{HTML}{ffd7d7}
\definecolor{ffb0b0}{HTML}{ffb0b0}
\definecolor{ff4a4a}{HTML}{ff4a4a}
\definecolor{ff2525}{HTML}{ff2525}
\definecolor{ff6161}{HTML}{ff6161}
\definecolor{ff4444}{HTML}{ff4444}
\definecolor{ff7b7b}{HTML}{ff7b7b}
\definecolor{ffe3e3}{HTML}{ffe3e3}
\definecolor{ff7676}{HTML}{ff7676}
\definecolor{fff5f5}{HTML}{fff5f5}
\definecolor{ff7d7d}{HTML}{ff7d7d}
\definecolor{ffb7b7}{HTML}{ffb7b7}
\definecolor{ff2d2d}{HTML}{ff2d2d}
\definecolor{ff1a1a}{HTML}{ff1a1a}
\definecolor{ff2929}{HTML}{ff2929}
\definecolor{ff8f8f}{HTML}{ff8f8f}
\definecolor{ff5353}{HTML}{ff5353}
\definecolor{ffe8e8}{HTML}{ffe8e8}
\definecolor{ff1c1c}{HTML}{ff1c1c}
\definecolor{ff2f2f}{HTML}{ff2f2f}
\definecolor{ff7070}{HTML}{ff7070}
\definecolor{ff1e1e}{HTML}{ff1e1e}
\definecolor{ff0303}{HTML}{ff0303}
\definecolor{ff6767}{HTML}{ff6767}
\definecolor{ff0e0e}{HTML}{ff0e0e}
\definecolor{ffbfbf}{HTML}{ffbfbf}
\definecolor{ff2b2b}{HTML}{ff2b2b}
\definecolor{ffeeee}{HTML}{ffeeee}
\definecolor{ff1515}{HTML}{ff1515}
\definecolor{ffd6d6}{HTML}{ffd6d6}
\definecolor{ff8888}{HTML}{ff8888}
\definecolor{ffeaea}{HTML}{ffeaea}
\definecolor{ff9393}{HTML}{ff9393}
\definecolor{ffa2a2}{HTML}{ffa2a2}
\definecolor{ff9b9b}{HTML}{ff9b9b}
\definecolor{ff2020}{HTML}{ff2020}
\definecolor{ff8e8e}{HTML}{ff8e8e}
\definecolor{ff4242}{HTML}{ff4242}
\definecolor{ff8b8b}{HTML}{ff8b8b}
\definecolor{ff8787}{HTML}{ff8787}
\definecolor{ff4040}{HTML}{ff4040}
\definecolor{b2b2b2}{HTML}{b2b2b2}
\definecolor{cccccc}{HTML}{cccccc}
\definecolor{e5e5e5}{HTML}{e5e5e5}
\begin{document}

\title{Novel Features for Time Series Analysis:\\A Complex Networks Approach} 
\author{Vanessa Freitas Silva \and
        Maria Eduarda Silva \and 
        Pedro Ribeiro \and
        Fernando Silva
}
\institute{Vanessa Freitas Silva \at
              CRACS-INESC TEC, Faculdade de Ciências, Universidade do Porto, Porto, Portugal \\
              \email{vanessa.silva@dcc.fc.up.pt} \\
              \emph{orcid.org/0000-0001-9828-0757}
           \and
           Maria Eduarda Silva \at
              LIAAD-INESC TEC, Faculdade de Economia, Universidade do Porto, Porto,Portugal \\
              \email{mesilva@fep.up.pt} \\
              \emph{orcid.org/0000-0003-2972-2050}
            \and
            Pedro Ribeiro \at
              CRACS-INESC TEC, Faculdade de Ciências, Universidade do Porto, Porto, Portugal \\
              \email{pribeiro@fc.up.pt} \\
              \emph{orcid.org/0000-0002-5768-1383}
            \and
            Fernando Silva \at
              CRACS-INESC TEC, Faculdade de Ciências, Universidade do Porto, Porto, Portugal \\
              \email{fmsilva@fc.up.pt} \\
              \emph{orcid.org/0000-0001-8411-7094}
}
\date{Received: date / Accepted: date}
\maketitle

%%%%%%%%%%%%%%%%%%%%%%%%%%%%%%%%%%%%%%%%%%%%%%%%%%%%%%%%%%%%%%%%%%%

\begin{abstract}
Being able to capture the characteristics of a time series with a feature vector is a very important task with a multitude of applications, such as classification, clustering or forecasting. Usually, the features are obtained from linear and nonlinear time series measures, that may present several data related drawbacks.
In this work we introduce \textit{NetF} as an alternative set of features, incorporating several representative topological measures of different complex networks mappings of the time series. Our approach does not require data preprocessing and is applicable regardless of any data characteristics. Exploring our novel feature vector, we are able to connect mapped network features to properties inherent in diversified time series models, showing that \textit{NetF} can be useful to characterize time data. 
Furthermore, we also demonstrate the applicability of our methodology in  clustering synthetic and benchmark time series sets, comparing its performance with more conventional features, showcasing how \textit{NetF} can achieve high-accuracy clusters. Our results are very promising, with network features from different mapping methods capturing different properties of the time series, adding a different and rich feature set to the literature. 

\end{abstract}

\section{Introduction}
\label{sec:intro}

Time series, which can be thought of as collections of observations indexed by time, are ubiquitous in all domains from climate studies or health monitoring to financial data analysis. There is a plethora of statistical models in the literature adequate to describe the behaviour of time series~\cite{Sumway2017}. 
However, technological developments, such as sensors and mobile devices, lead to  the gathering  of large amounts  of high  dimensional time indexed data for which appropriate methodological and computational tools are required.
With this purpose, recently, feature-based time series characterization  has become  a popular  approach among  time series data researchers~\cite{fulcher2018feature,henderson2021empirical,Wang2006} and proved useful 
for a wide range of temporal data mining tasks ranging from 
classification~\cite{fulcher2014highly}, 
clustering~\cite{Wang2006}, 
forecasting~\cite{montero2020fforma,talagala2018meta}, 
pattern detection~\cite{geurts2001pattern}, 
outlier or anomalies detection~\cite{hyndman2015large}, 
motif discovery~\cite{chiu2003probabilistic}, 
visualization~\cite{kang2017visualising} and 
generation of new data~\cite{kang2020gratis}, among others.

The main idea behind feature-based approaches is to construct feature vectors that aim to represent specific properties of the time series data by characterizing the underlying dynamic processes~\cite{fulcher2018feature,fulcher2017hctsa}. 
The usual methodologies for calculating time series features include concepts and methods from the linear time series analysis literature~\cite{Sumway2017}, such as autocorrelation, stationarity, seasonality and entropy, but also methods of nonlinear time-series analysis based on dynamic systems theory~\cite{fulcher2013highly,henderson2021empirical,Wang2006}. 
These methods usually involve parametric assumptions, parameter estimation, non-trivial calculations and approximations, as well as preprocessing tasks such as finding time series components, differencing and whitening thus  presenting drawbacks and computation issues related to the nature of the data, such as the length of the time series. 

This work contributes to the feature-based approach in time series analysis by proposing an alternative set of features based on complex networks concepts.

Complex networks describe a wide range of systems in nature and society by 
representing the existing interactions via graph structures~\cite{Barabasi2016}. 
Network science, the research area that studies complex networks, provides a vast set of topological graph measurements~\cite{Costa2007,peach2021hcga}, a well-defined set of problems such as community detection~\cite{Fortunato2017} or link prediction~\cite{Lu2011}, and a large track record of successful application of complex network methodologies to different fields~\cite{vespignani2018twenty}, 
including graph classification~\cite{bonner2016deep}. 

Motivated by the success of complex network methodologies and with the objective of acquiring new tools for the analysis of time series, several network-based  approaches have been recently proposed. These approaches involve mapping time series to the network domain. 
The mappings methods proposed in the literature may be divided into one of three categories depending on the underlying concept: proximity, visibility and transition~\cite{vanessa2020,zou2018complex}. 
Depending on the mapping method, the resulting networks capture specific properties of the time series. Some networks have as many nodes as the number of observations in the time series, as visibility graphs~\cite{Lacasa2008}, while others allow to reduce the dimensionality preserving the characteristics of the time dynamics, as the quantile graphs~\cite{Campanharo2011}. 
Network-based time series analysis techniques have been showing promising results and have been successful in the description, classification and clustering of time series. 
Examples include automatic classification of sleep stages~\cite{Zhu2014}, characterizing the dynamics of human heartbeat~\cite{Shao2010}, distinguishing healthy from non-healthy electroencephalographic series~\cite{Campanharo2017} and analysing seismic signals~\cite{Telesca2012}. 

\begin{figure}[hbt!]
	\centering 
	\includegraphics[scale=.34]{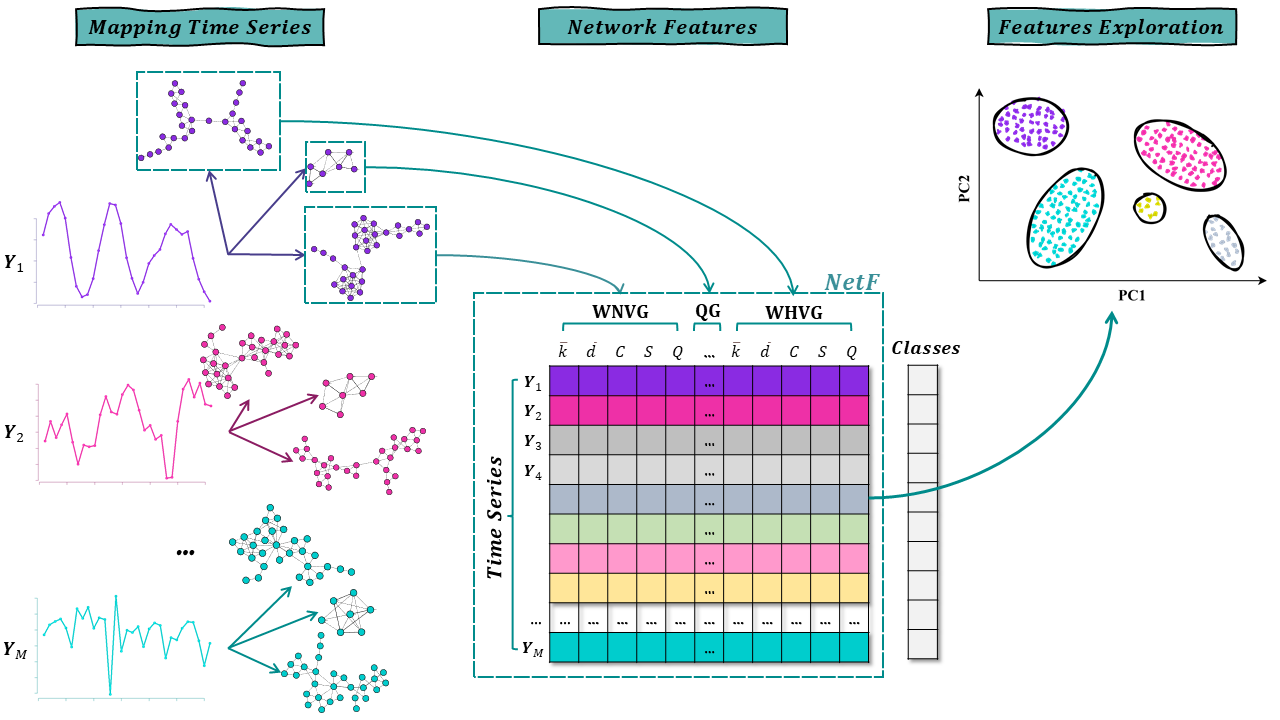}
	\caption{Schematic diagram of the network based features  approach to time series mining tasks.}
	\label{fig:overview}
\end{figure}

In this work we establish a new set of time series features, \textit{NetF}, by mapping  the time series into the complex networks domain. Further, we propose a  procedure for time series mining tasks  and  address the  question whether   time series  features based on complex networks are  a useful approach in time series mining tasks. Our proposed procedure, represented in Figure~\ref{fig:overview} comprises the following steps: map the time series into  (natural and horizontal) visibility  graphs and quantile graphs using appropriate mapping methods and compute  five specific topological measures for each network, thus establishing a vector of   15  features. These features are then used in mining tasks. The network topological metrics  selected, average weighted degree, average path length, number of communities, clustering coefficient and modularity, 
 measure global characteristics, are  simple to compute and to interpret in the graph context and commonly  used in network analysis, thus capable of providing useful information about the structure and properties of the underlying systems.  

To investigate the relevance of the set of features \textit{NetF} we analyse synthetic time series generated from a set of Data Generating Processes with a range of different characteristics. Additionally we consider  the problem of time series clustering from a feature-based  approach in synthetic, benchmark and a new time series data sets. \textit{NetF} features are assessed against two other sets of features, \textit{tsfeatures} and \textit{catch22}~\cite{tsfeatures,kang2017visualising,lubba2019catch22}. 
The results show that network science measures are able to capture  and preserve the  characteristics of the  time series data. 
We show that different topological measures from different mapping methods capture different characteristics of the data, complementing each other and providing new information when combined, rather than considered by themselves as is common in the literature. 
Clustering results of empirical data are balanced when compared to conventional approaches, 
in some data sets the proposed approach obtains better results, and in other data sets the results are quite similar between the approaches. 
The proposed set of features have the advantage of being always computable, which is not always possible using classical time series features. 

\textit{NetF}, the empirical study implementation, and the data sets presented here are available on GitHub\footnote{\url{https://github.com/vanessa-silva/NetF}}.

We have organized this paper as follows. 
Section~\ref{sec:backg} introduces basic concepts of time series and complex networks, setting the notation for the remainder of the paper, and also presents the mapping methods and networks measurements used. 
Next, in Section~\ref{sec:nov_features} the novel features of the time series proposed in this work are presented  and a study of these features is carried out in order to characterize properties of the time series. 
In Section~\ref{sec:emp_exper} the time series clustering tasks are performed as an example of application of network-based features, 
synthetic and empirical data sets are used, they are briefly described and compared to two other classical time series approaches. 
The results corresponding to the three approaches are presented. 
Finally, Section~\ref{sec:concl} presents the conclusions and some comments.

\section{Background}
\label{sec:backg}

\subsection{Time Series}
\label{subsec:ts}

A time series $\boldsymbol{Y}=(Y_1, \ldots, Y_T)$ is a set of  observations  obtained over time, usually at equidistant time points. 
A time series differs from a random sample in that the observations are ordered  in time and usually present serial correlation that must be account for in all statistical and data mining tasks. 
Time series analysis refers to the collection of procedures developed to systematically solve the statistical problems posed by the serial correlation. 
Statistical time series analysis relies on a set of concepts, measures and models designed to capture the essential characteristics of the data, namely, trend, seasonality, periodicity, autocorrelation, skewness, kurtosis and heteroscedasticity~\cite{Wang2006}. 
Other  concepts like self-similarity, non-linearity structure and chaos, stemming from non-linear science are also used to characterize time series~\cite{bradley2015nonlinear}. 

Several classes of  statistical models that provide plausible descriptions of the characteristics of the time series data have been developed with a view to forecasting and simulation~\cite{Box2015}. 
The statistical models for time series may be broadly classified as \textit{linear} and \textit{nonlinear}, referring usually to the functional forms of  conditional mean and variance. 
\textit{Linear time series models} are models for which the  conditional mean is a linear function of past values of the time series.  
The most popular class are the AutoRegressive Moving Average (ARMA) models. As particular cases of ARMA models we have:  the white noise (WN), a sequence of independent and identically distributed observations, the AutoRegression (AR) models, which specify a linear relationship between the current and past values and the Moving Average (MA) models, which specify the linear relationship between the current value and past stochastic terms.  ARMA models have been extended to incorporate non-stationarity (unit root) ARIMA models and long memory characteristics, ARFIMA models. 
Many time series data present characteristics that cannot be represented by linear models such as volatility, asymmetry, different regimes and clustering effects. To model these effects, non-linear specifications for the conditional mean and for the conditional variance  lead to different classes of \textit{nonlinear time series models}, such as Generalized AutoRegressive Conditional Heteroskedastic (GARCH) type models specified by the conditional variance and developed mainly for financial time series, 
threshold models and Hidden Markov models that allow for different regimes and models for integer valued time series, INAR. 
Definitions, properties and details about these models are given in Appendix~\ref{app:TSmodels}.

\subsection{Complex Networks}
\label{subsec:cn}

\textit{Graphs} are mathematical structures  appropriate for modelling complex systems which are characterized by a set of elements that interact with each other and exhibit collective properties~\cite{Costa2011}. 
Typically, graphs exhibit non-trivial topological properties due to the characteristics of the underlying system, and so they are often called complex networks. 

A graph (or network), $G$, is an ordered pair $(V, E)$, where $V$ represents the set of \textit{nodes} and $E$ the set of \textit{edges} between pairs of elements of $V$. 
Two nodes $v_{i}$ and $v_{j}$ are neighbours if they are connected by an edge $(v_{i},v_{j}) \in E$. 
The edges can be termed as \textit{directed}, if the edges connect a source node to a target node, or \textit{undirected}, if there is no such concept of orientation. 
A graph is also termed \textit{weighted} if there is a weight, $w_{i,j},$ associated with the edge $(v_{i}, v_{j})$. 

Network science has served many scientific fields in problem solving and analyzing data that is directly or indirectly converted to networks. 
Currently, there is a vast literature on problems and successful applications~\cite{vespignani2018twenty}, as well as an extensive set of measurements of topological, statistical, spectral and combinatorial properties of networks~\cite{Albert2002,Barabasi2016,Costa2007,peach2021hcga}, capable of differentiating particular characteristics of the network data. 
Examples include measures of node centrality~\cite{oldham2019consistency}, graph distances~\cite{li2021analyzing}, clustering and community~\cite{malliaros2013clustering}, among an infinity of them. 
Many of these topological measurements involve the concepts of \textit{paths} and graph \textit{connectivity}. 
A path is a sequence of distinct edges that connect consecutive pairs of nodes. 
And, consequently, two nodes are said to be connected if there is a path between them and disconnected if no such path exists. 
Thus, some measurements are based on the length (number of edges) of such connecting paths~\cite{Costa2007}.

\subsection{Mapping Time Series into Complex Networks} 
\label{sec:mapp}

In the last decade several network-based time series analysis approaches have been proposed. 
These approaches are based on mapping the time series into the network domain. 
The mappings proposed in the literature are essentially based on concepts of visibility, transition probability and proximity~\cite{vanessa2020,zou2018complex}. 
In this work we  use  visibility graph and quantile graph methods which are based on  visibility and transition probability concepts, respectively.  
Next, we briefly describe these methods.

\subsubsection{Visibility Graphs}
\label{subsec:vg}

Visibility graphs (VG) establish connections (edges) between the time stamps  (nodes) using visibility lines between the  observations, where nodes are associated with the natural ordering of observations.  
Two variants of this method are as follows.

The \textit{Natural Visibility Graph} (NVG)~\cite{Lacasa2008} is based on the idea that each observation, $Y_t$, of the time series is seen as a vertical bar with height equal to its numerical value and that these bars are laid in a landscape where the top of a bar is visible from the tops of other bars. 
Each time stamp, $t$, 
is mapped into a node in the graph and the edges $(v_i, v_j)$, for $i,j = 1 \ldots T$, $i \neq j$, are established if there is a line of visibility between the corresponding data bars that is not intercepted. 
Formally, two nodes $v_{i}$ and $v_{j}$ are connected if any other observation $(t_{k}, Y_{k})$ with $t_{i}<t_{k}<t_{j}$ satisfies the equation:
\begin{eqnarray}
	Y_{k} < Y_{j}+(Y_{i}-Y_{j})\frac{(t_{j}-t_{k})}{(t_{j}-t_{i})}.
\end{eqnarray}

\begin{figure}[hbt!]
    \centering
    \subfloat[Toy time series] {
        \includegraphics[scale = 0.5,keepaspectratio]{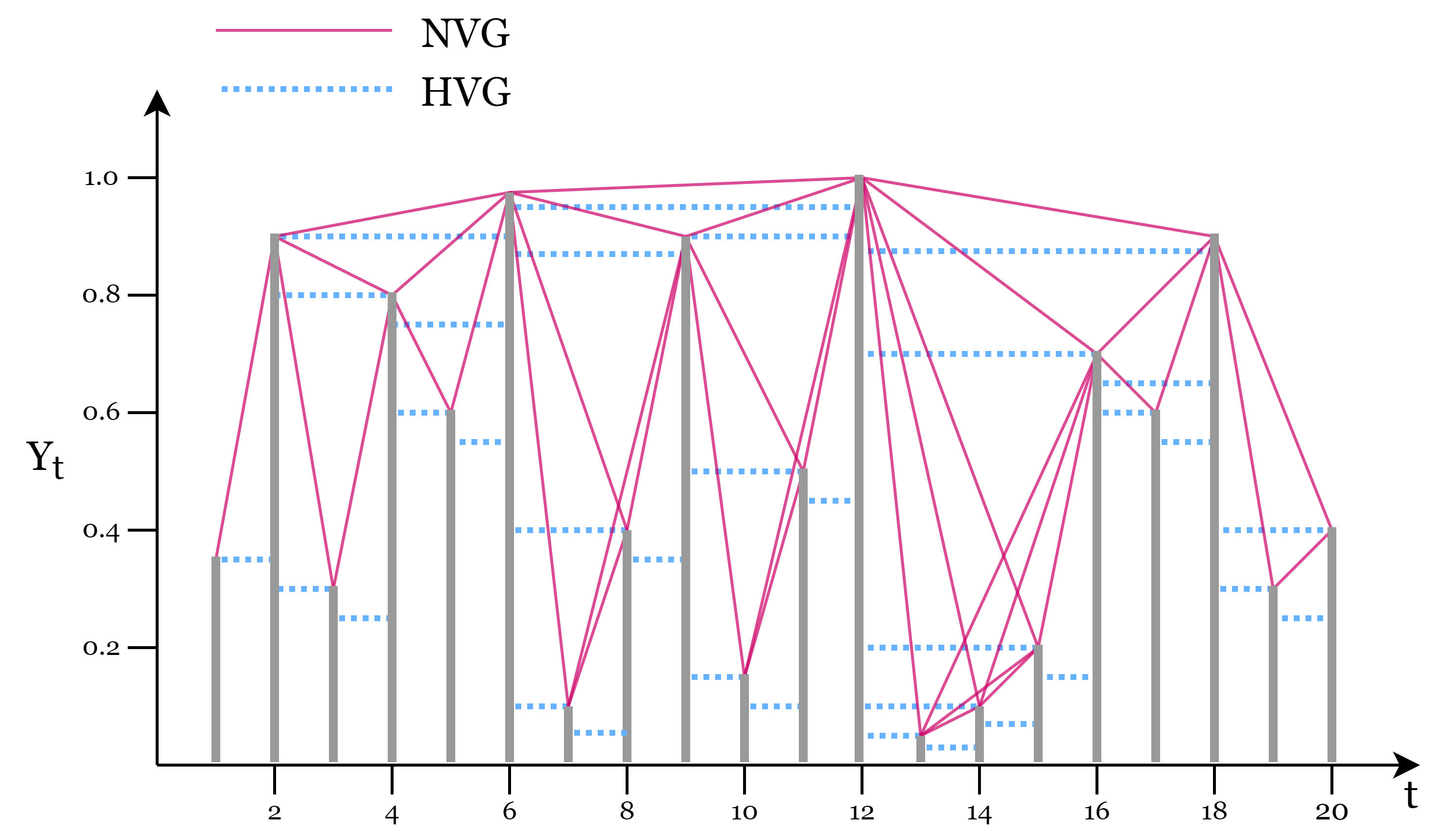}
		\label{Figure_1a}
    }
    \quad
    \subfloat[NVG and HVG]{
		\includegraphics[scale = 0.21,keepaspectratio]{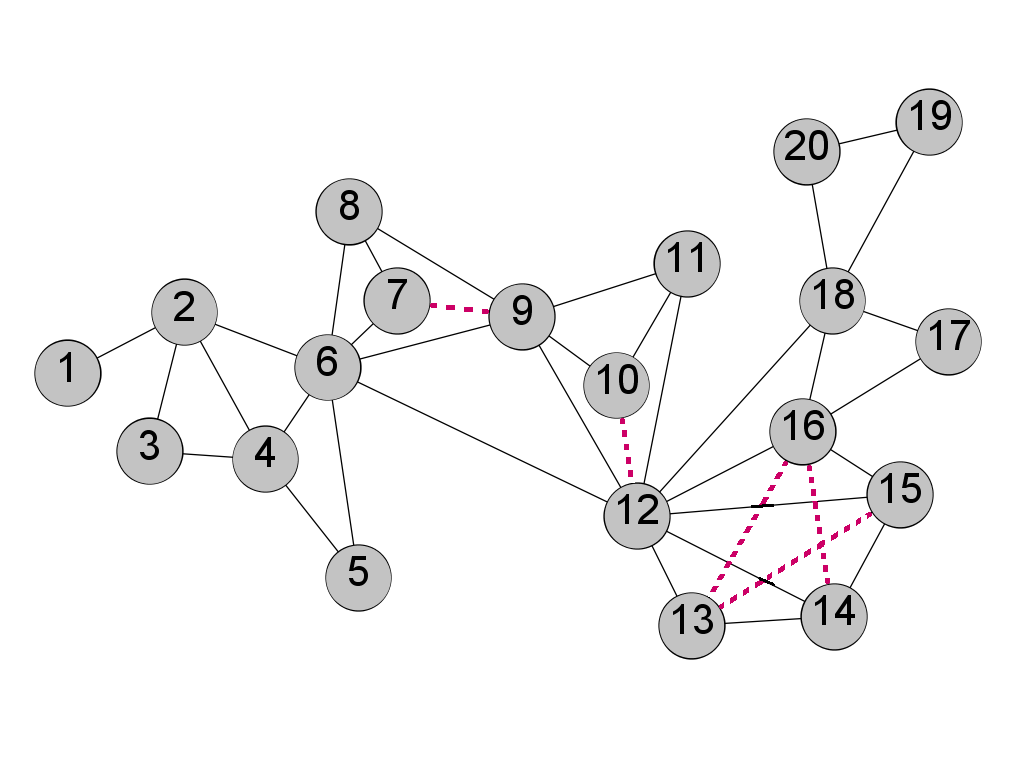}
		\label{Figure_1b}
	}
    \caption{Illustrative example of the two visibility graph algorithms. 
    (a) toy time series and corresponding visibility lines between data bars. Solid pink lines represent the natural visibility lines corresponding to the NVG method, and dashed blue lines represent the horizontal visibility lines corresponding to the HVG method. 
    (b) network generated by the corresponding mappings. 
    The NVG is the graph with all edges, including the edges represented by the dashed pink lines, and the HVG is the subgraph that does not include these edges.\\
    \textit{Source}: Modified from~\cite{vanessa2020}.}
\label{Figure_1}
\end{figure}

We give a simple illustration of this algorithm in Figure~\ref{Figure_1}. 
\vspace{-0.4cm}

\vspace{-0.4cm}
NVGs are always connected, each node $v_i$ sees at least its neighbors $v_{i-1}$ and $v_{i+1},$ and are always undirected unless we consider the direction of the time axis~\cite{vanessa2020}.  
The network is also invariant under affine transformations of the data~\cite{Lacasa2008} because the visibility criterion is invariant under rescheduling of both the horizontal and vertical axis, as well as in vector translations, that is, each transformation $\boldsymbol{Y}'= a\boldsymbol{Y} + b,$ for $a \in \mathbb{R}$ and $b \in \mathbb{R},$ leads to the same NVGs~\cite{vanessa2020}.

Eventual sensitivity of NVGs  to noise is attenuated by assigning a weight to the edges. 
Define $ w_{i,j} = 1 / \sqrt{(t_j-t_i)^2 + (Y_j - Y_i)^2}$ the weight associated  with the edge $(v_i, v_j)$~\cite{bianchi2017multiplex}. 
This weight is related to the Euclidean distance between the points $(t_i, Y_i)$ and $(t_j, Y_j).$ 
Thus, the resulting network from \textit{weighted} NVG (WNVG) method is a weighted  and undirected graph. 

The \textit{Horizontal Visibility Graph} (HVG)~\cite{Luque2009} is a simplification of the NVG method whose construction differs in the visibility definition: the visibility lines are only horizontal (see Figure~\ref{Figure_1}). 
Two nodes $v_{i}$ and $v_{j}$ are connected if, for all $(t_{k},Y_{k})$ with $t_{i} < t_{k} < t_{j}$, the following condition is met: 
\begin{eqnarray}
	Y_{i}, Y_{j} > Y_{k}.
\end{eqnarray}

Given a time series, its HVG is always a subgraph of its NVG. 
This is illustrated in Figure~\ref{Figure_1b} where all edges present in the HVG are also present in the NVG, but the converse is not true, the edges represented by dashed pink lines. 
HVG nodes will always have a degree less than or equal to that of  the corresponding NVG nodes. Therefore, there is some loss of quantitative information in HVG in comparison with NVG~\cite{Luque2009}. 
However, in terms of qualitative characteristics, the graphs preserve part of the data information, namely, the local information (the closest time stamps)~\cite{vanessa2020}. 

In a similar way to WNVG, we can assign weights to the edges of the HVG, $w_{i,j} = 1 / \sqrt{(t_j-t_i)^2 + (Y_j - Y_i)^2}$, resulting in a \textit{weighted} HVG (WHVG).

\subsubsection{Quantile Graphs}
\label{subsec:qg}

\begin{figure}[hbt!]
    \centering
    \subfloat[Toy time series] {
        \includegraphics[scale = 0.5,keepaspectratio]{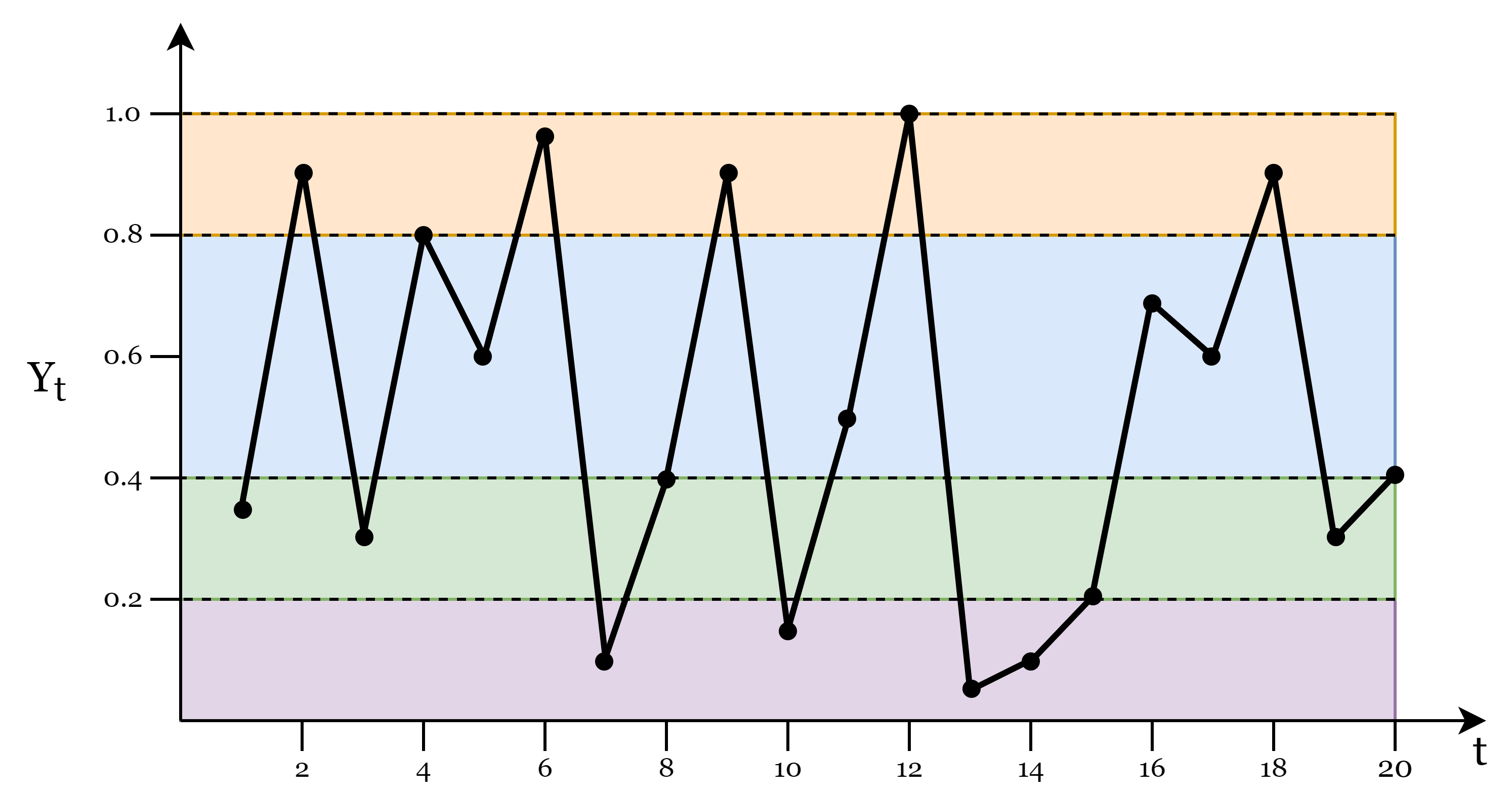}
    }
    \quad
    \quad
    \quad
    \subfloat[QG]{
		\includegraphics[scale = 0.5,keepaspectratio]{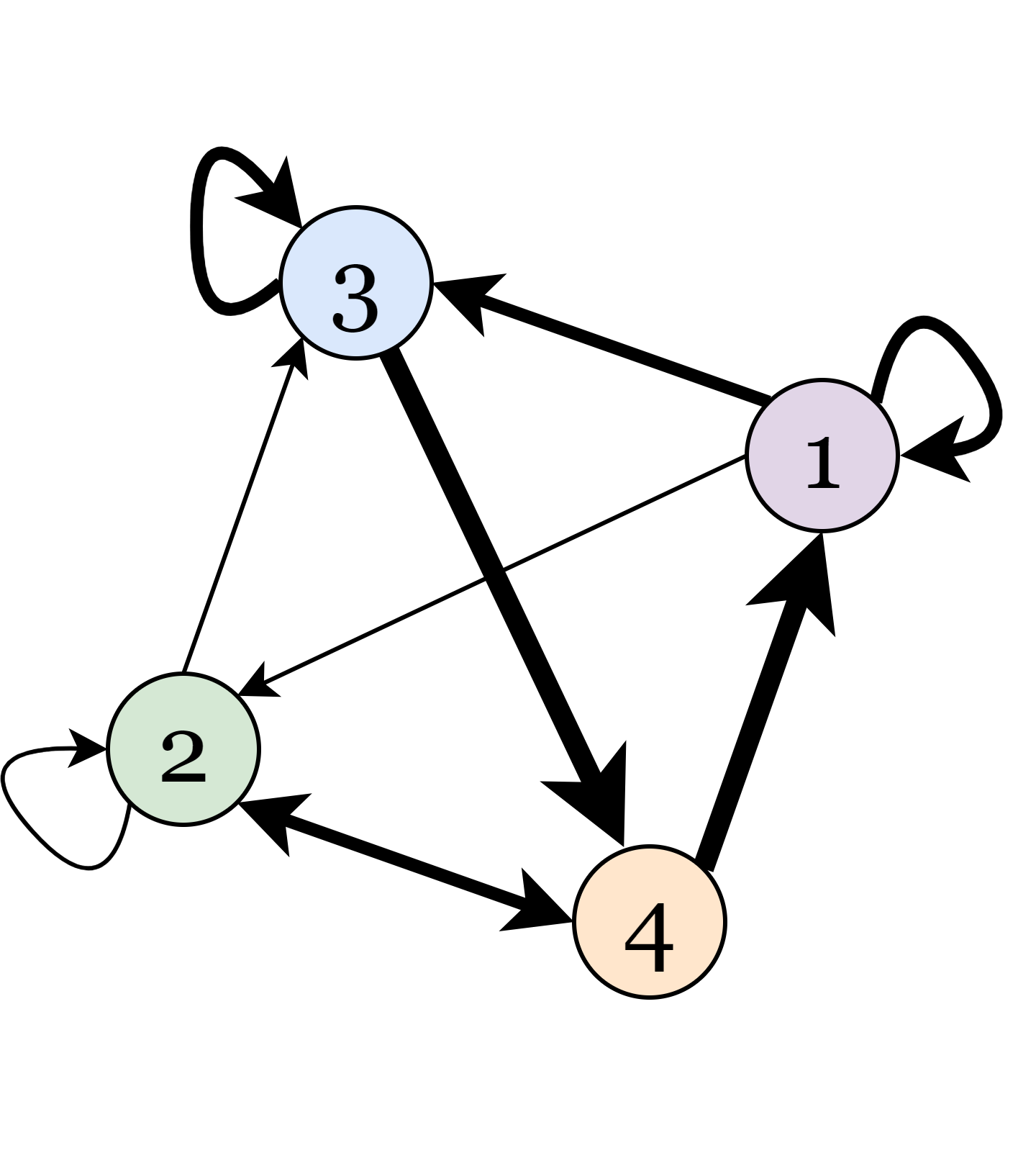}
	}
    \caption{Illustrative example of the quantile graph algorithm for $\eta = 4$. 
    (a) toy time series with coloured regions representing the different $\eta$ sample quantiles;
    (b)  network generated by the QG algorithm. 
    Repeated transitions between quantiles result in edges with larger weights represented by thicker lines.\\
    \textit{Source}: Reproduced from~\cite{vanessa2020}.}
\label{Figure_2}
\end{figure}

\textit{Quantile Graph} (QG)~\cite{Campanharo2011} are obtained from a mapping  based on transition probabilities. 
The method consists in assigning  the time series observations to bins that are defined by $\eta$ sample quantiles, $q_{1}, q_{2}, ..., q_{\eta}$.  
Each sample quantile, $q_{i}$, is mapped to a node $v_{i}$ of the graph and the edges between two nodes $v_{i}$ and $v_{j}$ are directed and weighted, $(v_{i}, v_{j}, w_{i,j})$, where $w_{i,j}$ represents the transition probability between quantile ranges. 
The adjacency matrix is a Markov transition matrix: $\sum_{j=1}^\eta w_{i,j} = 1$, for each $i = 1, \ldots, \eta,$ and the network is weighted, directed and contains self-loops\footnote{ A self-loop is an edge that connects a node to itself.}. 
We illustrate this mapping method in Figure~\ref{Figure_2}. 

The number of quantiles is usually much less than the length of the time series ($\eta \ll T$). 
If $\eta$ is too large the resulting graph  may not be connected, having isolated nodes\footnote{An isolated node is an node that is not connected by an edge to any other node.}, and if it is too small the QG may present  a significant loss of information, represented by large weights assigned to self-loops. 
The causal relationships contained in the process dynamics are captured by the connectivity of the QG.

\subsection{Complex Networks Topological Measures}
\label{subsec:cnm}

Complex networks have specific topological features which characterize the connectivity between their nodes and, consequently, are somehow reflected in the measurement processes~\cite{Costa2007}. 
There is a wide range of network topology measures capable of expressing the most relevant features of a network. They include global network measurements, which measure global properties involving all elements of the network, node-level or edge-level measurements, which measure a given feature corresponding to the nodes or edges, and "intermediate" measurements, which measure features of subgraphs in the network. 

Properties of centrality, distance, community detection and connectivity are central to understanding features of network structures. 
Centrality measures aim to quantify the importance of nodes and edges in the network depending on their connection topology. 
Path-based measures refer to sequences of edges that connect pairs of nodes, depend on the overall network structure and are useful for measuring network efficiency and information propagation capability. 
Communities and node connectivity are also very relevant features of networks, which measure how and which groups of nodes are better connected, measuring the clustering and resilience of the network. 

In this work we propose to study the average weighted degree, $\bar{k}$, average path length, $\bar{d}$, global clustering coefficient, $C$, number of communities, $S$, and modularity, $Q$, representing global measures of the features described above. 

The \textit{degree}, $k_{i}$, of a node $v_{i}$ represents the number of edges of $v_{i}$. It is a fairly important property that shows the intensity of connectivity in the node neighbourhood. 
In directed graphs we distinguish between \textit{in-degree}, $k_{i}^{in}$, the number of edges that point to $v_{i}$, and \textit{out-degree}, $k_{i}^{out}$, the number of edges that point from $v_{i}$ to other nodes. The total degree is given by $k_{i} = k_{i}^{in} + k_{i}^{out}$. 
For weighted graphs, we can calculate the \textit{weighted degree} by adding the edge weights instead of the number of edges. 
\textit{Average path length}, $\bar{d}$, is the arithmetic mean of the shortest paths, $d_{i,j}$, among all pairs of nodes, where the path length is the number of edges, or the sum of the edges weights for weighted graphs, in the path. It is a good measure of the efficiency of information flow on a network. 
The \textit{global clustering coefficient}, $C$, also known as \textit{transitivity}, is a measure which captures the degree to which the nodes of a graph tend to cluster, that is, the probability that two nodes connected to a given node are also connected. 
In this work, we refer to network communities as the grouping of nodes (potentially overlapping) that are densely connected internally and that can also be considered as a group of nodes that share common or similar characteristics. The \textit{number of communities}, $S$, is the amount of these groups on the network. 
The \textit{modularity}, $Q$, measures how good a specific division of the graph is into communities, that is, how different are the different nodes belonging to different communities. 
A high value of $Q$ indicates a graph with a dense internal community structure and sparse connections between nodes of different communities.

\section{\textit{NetF}: A Novel Set of Time Series Features}
\label{sec:nov_features}

Over the last decades, several techniques for extracting time series features have been developed (see~\cite{barandas2020tsfel,christ2018time,fulcher2017hctsa,fulcher2013highly,tsfeatures,lubba2019catch22,feasts} for more details). 
Most of these techniques have in common the definition of a finite set of statistical features, such as autocorrelation, existence of unit roots, periodicity, nonlinearity, volatility among others, to capture the global nature of the time series. 

In this work we introduce \textbf{\textit{NetF}} as an alternative set of features. Our approach differs from those previously mentioned in that we leverage the usage of different complex network mappings to offer a set of time series features based on the topology of those networks.  
One of the main advantages of this approach comes from the fact that the mapping methods (Section~\ref{sec:mapp}) do not require typical time series preprocessing tasks, such as decomposing, differencing or whitening. Moreover, our methodology is applicable to any time series, regardless of its characteristics.

\subsection{The 15 Features of \textit{NetF}}

\begin{figure}[hbt!]
	\centering 
	\includegraphics[scale=.38]{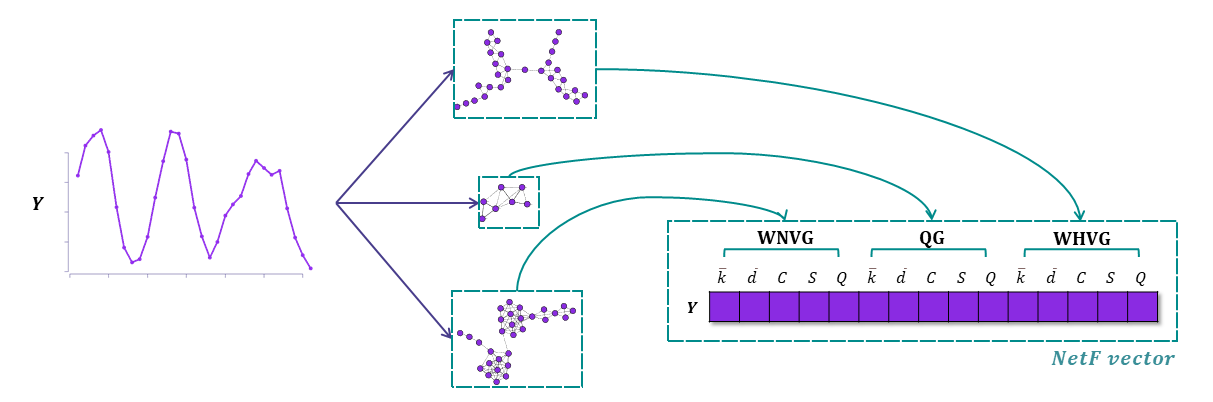}
	\caption{Schematic diagram of the \textit{NetF} set of features. A time series $\mathbf{Y}$ is mapped into three complex networks (WNVG, WHVG and QG) and for each of these networks five topological measures are taken ($\bar{k}$, $\bar{d}$, $C,$ $S$ and $Q$), resulting in the \textit{NetF} vector containing 15 features. 
	}
	\label{fig:over_NetF}
\end{figure}

\textit{NetF} is constituted by 15 different features, as depicted in Figure~\ref{fig:over_NetF}. These features correspond to the concatenation of five different topological measures, as explained in Section~\ref{subsec:cnm} ($\bar{k}$, the average weighted degree; $\bar{d}$, the average path length; $C$, the clustering coefficient; $S$, the number of communities; $Q$, the modularity), each of them applied to three different mappings of the time series, as explained in Section~\ref{sec:mapp} (\textit{WNVG}, the weighted natural visibility graph; \textit{WHVG}, the weighted horizontal visibility graph; \textit{QG}, the quantile graph). 
\vspace{-0.4cm}

\vspace{-0.4cm}
Our main goal is to provide a varied set of representative features that expose different properties captured by the topology of the mapped networks, providing a rich characterization of the underlying time series. 

The topological features themselves were selected so that they represent global measures of centrality, distance, community detection and connectivity, while still being accessible, easy to compute and widely used in the network analysis domain.

\subsection{Implementation Details}

Conceptually, \textit{NetF} does not depend on the actual details of how it is computed. Nevertheless, with the intention of both showing the practicality of our approach, as well as providing the reader with the ability to reproduce our results, we now describe in detail how we computed the \textit{NetF} set of features in the context of this work.

To compute the WNVGs we implement the  \textit{divide \& conquer} algorithm proposed in~\cite{Lan2015} and for the WHVGs the algorithm proposed in~\cite{Luque2009}\footnote{\url{https://sites.google.com/view/lucaslacasa/research-topics/visibility-graphs\#h.27cb86b8e1ba43fd_148}}. To both we added the weighted version mentioned in Section~\ref{subsec:vg}, adding the respective weights to the edges. 
For the QGs we chose $ \eta = 50 $ quantiles,  as in~\cite{campanharo2016hurst}, and we implemented the method described in Section~\ref{subsec:qg} to create the nodes and edges of the networks. 
We used the sample quantile method, which uses a scheme of linear interpolation of the empirical distribution function~\cite{hyndman1996sample}, to calculate the sample quantiles (nodes) in support of the time series. 
To save the network structure as a graph structure, we used the \texttt{igraph}~\cite{igraph} package which also allows us to calculate the topological measures. 
Next, we briefly describe the methods and algorithms used by the functions we used to calculate the measures. 

The \textbf{average weighted degree ($\bar{k}$)} is calculated by the arithmetic mean of the weighted degrees $k_i$ of all nodes in the graph. 
In this work, the \textbf{average path length ($\bar{d}$)} follows an algorithm that does not consider edge weights, and use the breadth-first search algorithm to calculate the shortest paths $d_{i,j}$ between all pairs of vertices, both ways for directed graphs. 
For calculate the \textbf{clustering coefficient ($C$)}, the function that we use in this work ignores the edge direction for directed graphs. For this reason, before we calculate $C$ for QGs, which are directed graphs, we first transform them into an undirected graph, where for each pair of nodes which are connected with at least one directed edge the edge is converted to an undirected edge. And then, the $C$ is calculated by the ratio of the total number of closed triangles\footnote{A triangle is a set of three nodes with edges between each pair of nodes.} in the graph to the number of triplets\footnote{A triplet is a set of three nodes with at least edges between two pairs of nodes.}. 
The function we use in this work to calculate the \textbf{number of communities ($S$)} in a network, calculates densely connected subgraphs via random walks, such that short random walks tend to stay in the same community. See the Walktrap community finding algorithm~\cite{pons2006computing} for more details. 
And to calculate the \textbf{modularity ($Q$)} of a graph in relation to some division of nodes into communities we measure how separated are the nodes belonging to the different communities are as follows:
\begin{equation*}
    Q = \frac{1}{2|E|} \sum_{i,j} \left[ w_{i,j} - \frac{k_i k_j}{2|E|} \right] \delta \left( c_i, c_j \right),
\end{equation*}
where $|E|$ is the number of edges, $c_i$ and $c_j$ the communities of $v_i$ and $v_j$, respectively, and $\delta(c_i, c_j) = 1$ if $v_i$ and $v_j$ belong to the same community ($c_i = c_j$) and $\delta(c_i, c_j) = 0$ otherwise.
We performed all implementations and computations in R~\cite{R}, version 4.0.3 and a set of packages. 

For reproducibility purposes, the source code and the datasets are made available in \url{https://github.com/vanessa-silva/NetF}. % and the datasets and results are made available in \url{https://www.dcc.fc.up.pt/~vanessa.silva/NetF}.

\subsection{Empirical Evaluation}
\label{subsec:charact_models}

In this section we investigate, via synthetic data sets,  whether the set of features introduced above are useful for characterizing time series data.

To this end, 
we consider a  set of eleven linear and nonlinear time series models, denoted by Data Generating Processes (DGP),   which present a wide range of  characteristics summarized in Table~\ref{table:1}. A detailed description of the DGP and computational details are given in Appendix~\ref{app:TSmodels}. 
For each of the DGP's in Table~\ref{table:1} we generated 100 realizations  of length $T = 10000.$ Following the steps presented in Figure~\ref{fig:overview}, we map each realization into three networks and extract the corresponding topological measures. 
The resulting time series features, organized by mapping,  are summarized, mean and standard deviation,  in Tables~\ref{app_table:1}  to~\ref{app_table:3}. Note that the values have been Min-Max normalized for comparison purposes since the range of the different features vary across models. 
\begin{table}[!ht]
\scriptsize
\centering
\caption{Summary about the data generating process (time series models)  of the synthetic data. Parameters, main characteristic of the data sets and notation is also included.   See Appendix~\ref{app:TSmodels} for more details.}
\begin{tabular}{lllc}
\hline
\textbf{Process} & \textbf{Parameters} & \textbf{Main Property} & \textbf{Notation} \\
\hline

\rule{0pt}{12pt}White Noise & $\epsilon_{t} \sim N(0,1)$ & Noise effect &  \texttt{WN} \\

\rule{0pt}{12pt}AR$(1)$ & $\phi_{1} \in \{-0.5, 0.5\}$ & Smoothness & \texttt{AR(1)-0.5} \\
 & & & \texttt{AR(1)0.5} \\

\rule{0pt}{12pt}AR$(2)$ & $\phi_{1} = 1.5$, $\phi_{2} = -0.75$ & Pseudo-periodic & \texttt{AR(2)} \\

\rule{0pt}{12pt}ARIMA$(1,1,0)$ & $\phi_{1} = 0.7$ & Stochastic trend & \texttt{ARIMA} \\

\rule{0pt}{12pt}ARFIMA$(1,0.4,0)$ & $\phi_{1} = 0.5$ & Long memory effect & \texttt{ARFIMA} \\

\rule{0pt}{12pt}SETAR$(1)$ & $\alpha = 0.5$, $\beta = -1.8$, $\gamma = 2$, & Regime-dependent  & \multirow{2}{*}{\texttt{SETAR}} \\
 & $r = -1$ & autocorrelation\footnote{The two regimes have quite different autocorrelation properties: in the first the correlation is positive while in the second alternates between positive and negative values.} & \\

\rule{0pt}{12pt}Poisson-HMM & $N = 2$, {$\bigl[ \begin{smallmatrix} 0.9 & 0.1\\ 0.1 & 0.9 \end{smallmatrix} \bigr]$} $\lambda \in \{10, 15\}$ & State transitions & \texttt{HMM} \\

\rule{0pt}{12pt}GARCH$(1,1)$ & $\omega = 10^{-6}$, $\alpha_1 = 0.1$, & Persistent periods of high & \multirow{2}{*}{\texttt{GARCH}} \\
 & $\beta_1 = 0.8$ & or low volatility & \\
 
\rule{0pt}{12pt}EGARCH$(1,1)$ & $\omega = \left(10^{-6} -0.1 \sqrt{2/\pi} \right)$, & Asymmetric effects of & \multirow{2}{*}{\texttt{EGARCH}} \\
 & $\alpha_1 = 0.1$, $\beta_1 = 0.01$, $\gamma_1 = 0.3$ & positive and negative shock & \\
 
\rule{0pt}{12pt}INAR$(1)$ & $\alpha = 0.5$, $\epsilon_{t} \sim Po(1)$ & Correlated counts & \texttt{INAR} \\

\hline
\end{tabular}
\label{table:1}
\end{table}

\newpage
%\vspace{-0.4cm}
\begin{figure}[hbt!]
	\centering 
	\includegraphics[scale=.45]{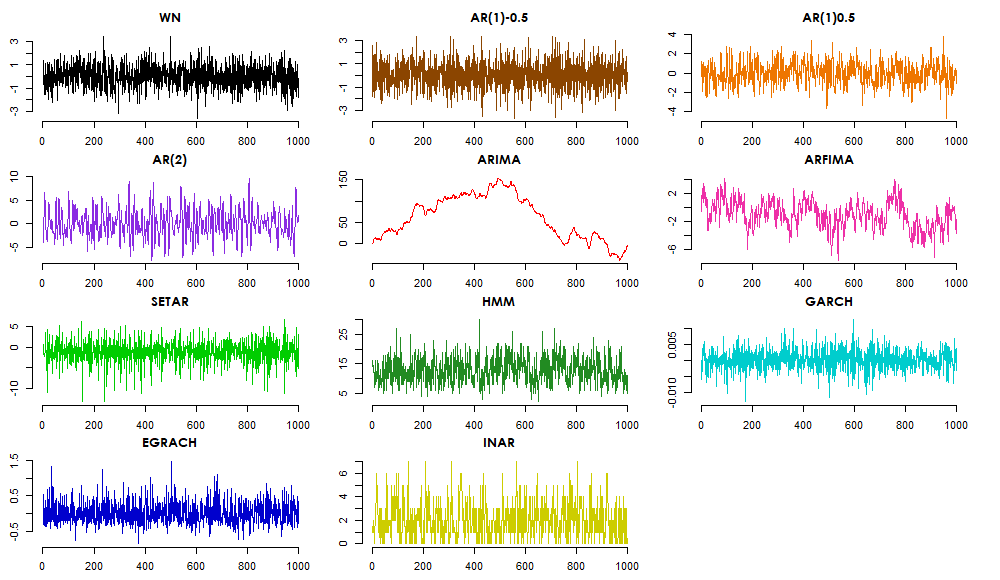}
	\caption{Plot of one instance of each simulated time series model.}
	\label{Figure_3}
\end{figure}

\vspace{-0.4cm}
WNVGs (Table~\ref{app_table:1}) present lowest values for the  clustering coefficient $(C)$ for ARIMA models. Models producing time series with more than one state (\texttt{HMM} and \texttt{SETAR}) present lower average weighted degree but higher number of communities $(S).$  The later values are comparable to those for  \texttt{AR(2)} time series, fact that can be explained by the pseudo-periodic nature of the particular \texttt{AR(2)} model entertained here.
WHVGs (Table~\ref{app_table:2}) present average weighted degree $(\bar{k})$ approximately 0 for   \texttt{HMM}'s  and  approximately 1 for  \texttt{GARCH} and \texttt{EGARCH}. This indicates that  \texttt{HMM} time series  have, on average, horizontal visibility for more distant points (in time and/or value), while  the opposite is true for heteroskedastic time series. 
The clustering coefficient $(C)$ is lowest (approximately 0) for networks obtained from \texttt{INAR}  time series, indicating that  most points  have visibility only for their two closest neighbors. 
QGs (Table~\ref{app_table:3})  present high values of average path length, $(\bar{d}),$ for \texttt{ARIMA}, contrasting with all other DGP which present low values. On the other hand,  the $(C)$ for \texttt{ARIMA} presents low values while all other DGP's present high values. 

\begin{figure}[hbt!]
	\centering 
	\includegraphics[scale=0.45]{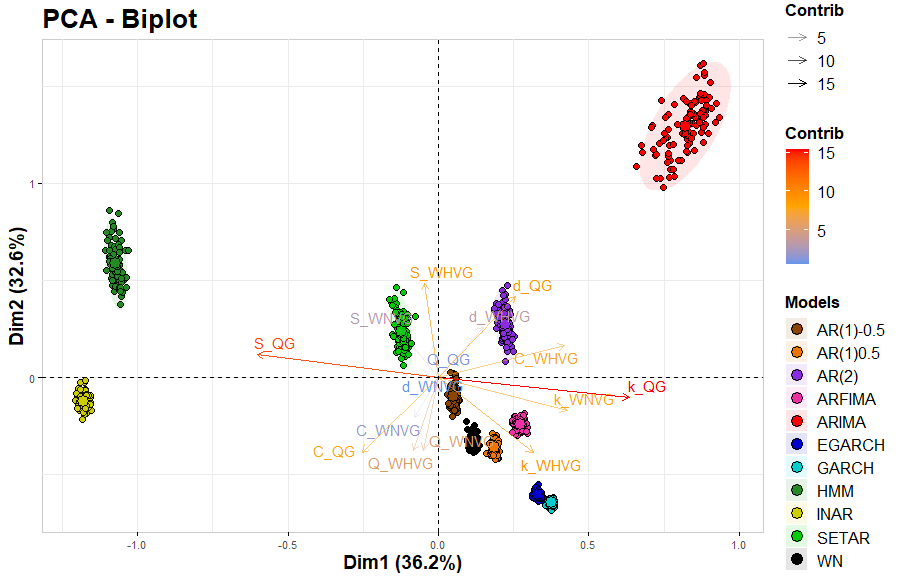}
	\caption{Bi-plot of the first two PC's for the synthetic data set. Each Data Generating Process (DGP) is represented by a color  and the arrows represent the contribution of the corresponding feature to the PC's:  the  larger the size, the sharper the color and the  closer to the red  the greater the contribution of the feature. 
	Features grouped together are positively correlated while those placed on opposite quadrants are negatively correlated.}
	\label{Figure_5}
\end{figure}

The next step is to study  the feature space to understand which network features capture specific properties of the time series models.  
Figure~\ref{Figure_5} represents a bi-plot obtained using the 15 features (5 for each mapping method) and with the two PC's explaining 68.8\% of the variance. 
It is noteworthy that the eleven groups of time series models are clearly identified and arranged in the bi-plot according to their main characteristics. 
Overall, we can say that the  number of communities of VGs, $S,$  are positively correlated among themselves and are negatively correlated with  the average weighted degree, $\bar{k},$ of \textit{NetF}. The average path length, $\bar{d},$ of WHVGs and QGs and the clustering coefficient, $C,$ of WHVG are positively correlated, but negatively to the $\bar{d}, C$ and $Q$ of WNVG, $Q$ of WHVG and $C$ of QGs. 
The features that most contribute to the total dimensions formed by the PCA are: $\bar{k}, S, Q$ and $\bar{d}$ of the QGs, $\bar{k}$ of the WNVGs, and $\bar{k}, S$ and $Q$ of the WHVGs (see Figure~\ref{app_fig2}). 

\newpage
The (stochastic) trend of  the \texttt{ARIMA}, in fact the only non-stationary DGP this data set, is represented by  high average path lengths, $\bar{d}, $ in WHVG and QG.
Discrete states in the data, \texttt{HMM,SETAR,INAR}, are associated with  the  number of communities, $S.$ 
The bi-plot further indicates that height average weighted degree, $\bar{k},$ mainly that of the WHVG, represents  heteroskedasticity in the time series, e.g., \texttt{GARCH} and \texttt{EGARCH}. Cycles, \texttt{AR(2)}, are captured by the clustering coefficient, $C.$
\vspace{-0.4cm}

\vspace{-0.4cm}
We also did an empirical study of the \textit{NetF} features in the context of clustering these synthetic data sets, and the results show that using the entire feature set leads to better performance than any possible subset. This showcases how the different features complement each other and how they capture different characteristics of the underlying time series. The details of this study can be seen in Appendix~\ref{app:tsmodels_clust}.

\section{Mining Time Series with \textit{NetF}}
\label{sec:emp_exper}

In this section we illustrate the usefulness of complex networks based time series features  in data mining tasks with a case study regarding time series clustering via feature-based approach~\cite{maharaj2019time}. Within this mining task we analyse the synthetic data set introduced in Section~\ref{subsec:charact_models}, benchmark empirical data sets from UEA \& UCR Time Series Classification Repository~\cite{UCRArchive}, the M3 competition data from package \texttt{Mcomp}~\cite{Mcomp}, the set "18Pairs" from package \texttt{TSclust}~\cite{TSclust}  and a new data set regarding the production of several crops across Brazil~\cite{ibge} using \textit{NetF} and two other sets of time series features, namely \textit{catch22}~\cite{lubba2019catch22} and \textit{tsfeatures}~\cite{Wang2006}, see Appendix~\ref{app:features}.

\subsection{Clustering Methodology} 
\label{subsec:descr}
\begin{figure}[hbt!]
	\centering 
	\includegraphics[scale=.45]{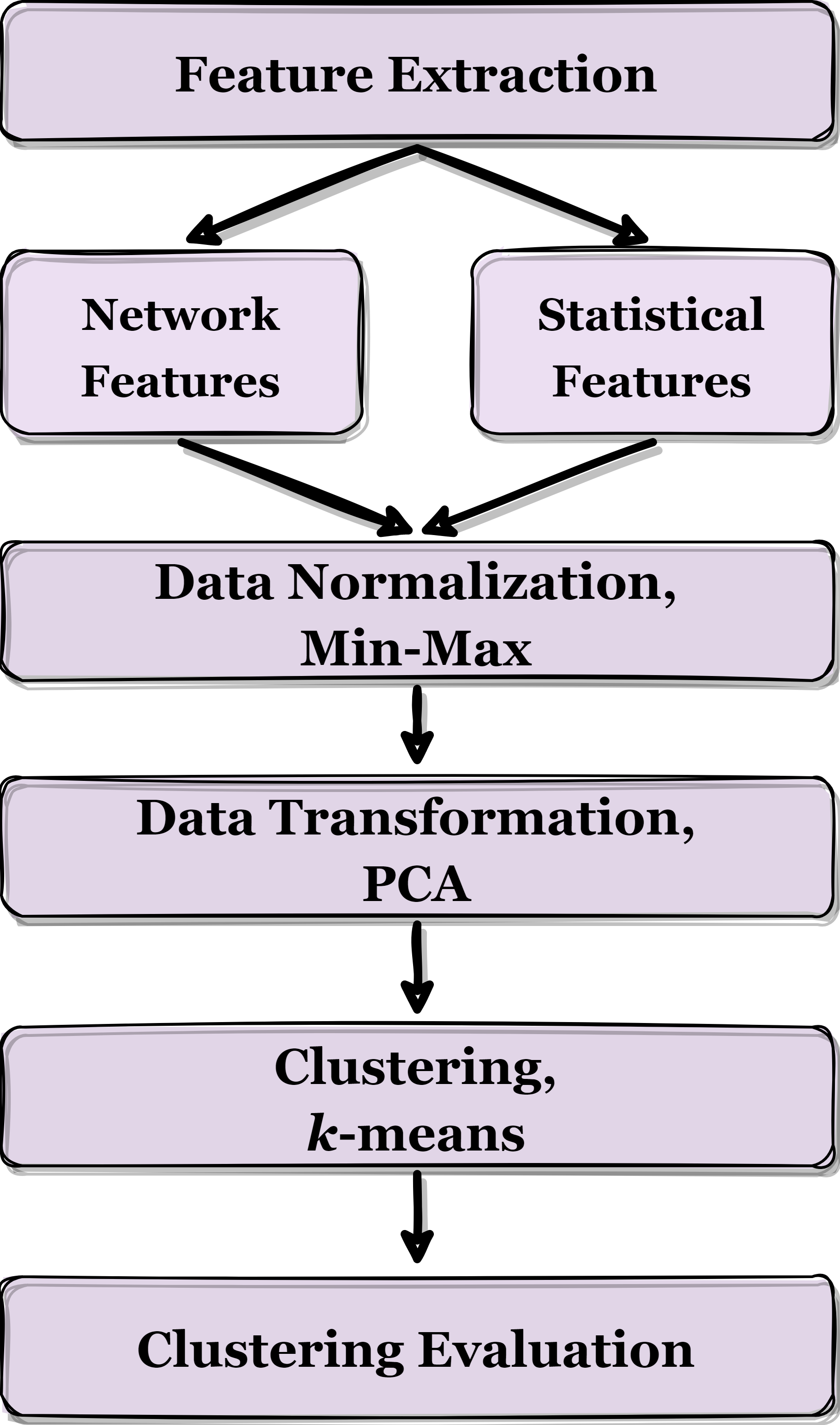}
	\caption{Schematic diagram for the time series clustering analysis procedure.}
	\label{fig:diagram}
\end{figure}

The overall procedure proposed here for feature-based clustering  is represented in Figure~\ref{fig:diagram}. 
\vspace{-0.5cm}

\vspace{-0.4cm}
Given a set of time series, compute the feature vectors which are then Min-Max rescaled into the $[0,1]$ interval and organized in a feature data matrix. Principal Components (PC) are computed (no need of z-score normalization within PCA) and finally a clustering algorithm is applied to the   PC's.  Among several algorithms available for clustering analysis, we opt for \textit{$k$-means}~\cite{jaaw28m} since it is fast and widely used for clustering.  Its main disadvantage is the need to pre-define the number of clusters. This issue will be discussed within each data set example. The clustering results are assessed using  appropriate evaluation metrics: \textit{Average Silhouette} (AS);   \textit{Adjusted Rand Index} (ARI) and  \textit{Normalized Mutual Information} (NMI) when the ground truth is available.  

\subsection{Data Sets and Experimental Setup}
\label{subsec:datasets}

We report the detailed results for the clustering exercise for the eleven  data sets summarized in Table~\ref{table:2}. The brief description of the data and  clustering results for the remaining data sets  is presented in Tables~\ref{app:table_utsc_res1}-\ref{app:table_utsc_res3}. 

The data sets in Table~\ref{table:2} belong to the UEA \& UCR Time Series Classification Repository~\cite{UCRArchive}, widely used in classification tasks,  the M3 competition data from package \texttt{Mcomp}~\cite{Mcomp} used for testing the performance of forecasting algorithms, the set "18Pairs", extracted from package \texttt{TSclust}~\cite{TSclust}  which represents pairs of time series of different domains. For all these we have true clusters and therefore clustering assessment measures ARI and NMI may be used. Additionally, we also analyse a set of observations comprising the production over forty three years of nine agriculture products in 108 meso-regions of Brazil~\cite{ibge}.  We note that the size of the ElectricDevices dataset, 16575 time series,  is different from the total available in the repository, as exactly 62 time series  return missing values for the entropy feature of the \textit{tsfeatures} set (see Appendix~\ref{app:features})  and so we decided to exclude these series from our analysis.

\newpage
\begin{table}[hbt!]
\centering

\caption{Brief description of the empirical time series data sets.}

\begin{tabular}{ |l|r|r|r|c| }
\hline
\multicolumn{1}{|c|}{\textbf{Data Set}} & \multicolumn{1}{c|}{\textbf{Size of}} & \multicolumn{1}{c|}{\textbf{Time Series}} & \multicolumn{1}{c|}{\textbf{Num. of}} & \multicolumn{1}{c|}{\textbf{Source}} \\
 & \multicolumn{1}{c|}{\textbf{Data Set}} & \multicolumn{1}{c|}{\textbf{Length}} & \multicolumn{1}{c|}{\textbf{Classes}} &  \\ 

\hline
{\scriptsize \textbf{18Pairs}} 				  & 36 	  & 1000	    & 18 &  \cite{TSclust} \\
\hline
{\scriptsize \textbf{M3 data}}	              & 3003  & $T \in [20, 144]$	& 6  &  \cite{Mcomp}  \\
\hline
{\scriptsize \textbf{CinC\_ECG\_torso}} 	  & 1420  & 1639	    & 4  &  \cite{UCRArchive} \\
\hline
{\scriptsize \textbf{Cricket\_X}} 			  & 780   & 300	        & 12  &  \cite{UCRArchive} \\
\hline
{\scriptsize \textbf{ECG5000}} 				  & 5000  & 140      	& 5  &  \cite{UCRArchive} \\
\hline
{\scriptsize \textbf{ElectricDevices}} 		  & 16575 & 96	        & 7  &  \cite{UCRArchive} \\
\hline
{\scriptsize \textbf{FaceAll}} 				  & 2250  & 131	        & 14 &  \cite{UCRArchive} \\
\hline
{\scriptsize \textbf{FordA}} 				  & 4921  & 500	        & 2  &  \cite{UCRArchive} \\
\hline
{\scriptsize \textbf{InsectWingbeatSound}} 	  & 2200  & 256     	& 11 &  \cite{UCRArchive} \\
\hline
{\scriptsize \textbf{UWaveGestureLibraryAll}} & 4478 & 945      	& 8  &  \cite{UCRArchive} \\
\hline
{\scriptsize \textbf{Production in Brazil}}	  & 108   & 198	        & 9  &  \cite{ibge} \\
\hline
\end{tabular}
\label{table:2}
\end{table}

\subsection{Results}

First,  we investigate the performance of \textit{NetF}, \textit{catch22} and \textit{tsfeatures} in the automatic determination of the number of clusters $k,$   using the clustering evaluation metrics, ARI, NMI and AS. The results  (see Table~\ref{app:table_bestK}) show overall similar values but we note that  \textit{NetF} seems to provide a value of  $k$  equal to or closer to the ground truth value (when available)  more often. For the Production in Brazil data, for which there is no ground truth, values for $k$ are obtained averaging 10 repetitions of the clustering procedure and using the silhouette method. The results of the 10 repetitions are represented in Figure~\ref{app:Figure_ASbestk} and   summarized in Table~\ref{table:5}.  

Next, fixing $k$ to the ground truth we perform the clustering procedure. The 
clustering evaluation metrics, mean over 10 repetitions,  are presented in Table~\ref{table:4}\footnote{The results for the remaining empirical time series data sets of the UEA \& UCR Time Series Classification Repository are presented in Tables~\ref{app:table_utsc_res1} to \ref{app:table_utsc_res3}.}.

\begin{table}[hbt!]
\centering
\caption{Clustering evaluation metrics obtained for the three approaches \textit{NetF}, \textit{tsfeatures} and \textit{catch22}. The values reflect the mean of 10 repetitions of the clustering analysis with number of classes equal to the ground truth (see Table~\ref{table:2}). 
The values in bold represent the best results.}
\begin{tabular}{ |l|c|c|c|c|c|c|c|c|c| }
\hline
 & \multicolumn{3}{c|}{\textbf{ARI}} & \multicolumn{3}{c|}{\textbf{NMI}} &  \multicolumn{3}{c|}{\textbf{AS}}\\
\multicolumn{1}{|c|}{\textbf{Data set}} & \multicolumn{3}{c|}{{\scriptsize $[-1,1]$}} & \multicolumn{3}{c|}{{\scriptsize $[0,1]$}} & \multicolumn{3}{c|}{{\scriptsize $[-1,1]$}} \\ 
\cline{2-10}
 & \multicolumn{1}{c|}{{\scriptsize \textit{tsf.}}} & \multicolumn{1}{c|}{{\scriptsize \textit{cat.}}} & \multicolumn{1}{c|}{{\scriptsize \textit{NetF}}} & \multicolumn{1}{c|}{{\scriptsize \textit{tsf.}}} & \multicolumn{1}{c|}{{\scriptsize \textit{cat.}}} & \multicolumn{1}{c|}{{\scriptsize \textit{NetF}}} & \multicolumn{1}{c|}{{\scriptsize \textit{tsf.}}} & \multicolumn{1}{c|}{{\scriptsize \textit{cat.}}} & \multicolumn{1}{c|}{{\scriptsize \textit{NetF}}} \\
\hline
{\tiny \textbf{18Pairs}} 				  & \textbf{0.51} & 0.39 & 0.49 & \textbf{0.89} & 0.86 & \textbf{0.89} & \textbf{0.42} & 0.32 & 0.34 \\

\hline
{\tiny \textbf{M3 data}} 				  & \textbf{0.14} & 0.13 & 0.13 & \textbf{0.21} & 0.19 & 0.18 & \textbf{0.36} & 0.22 & 0.31 \\

\hline
{\tiny \textbf{CinC\_ECG\_tors}} 	  & 0.31 & 0.32 & \textbf{0.45} & 0.37 & 0.35 & \textbf{0.52} & 0.23 & 0.19 & \textbf{0.31} \\
\hline
{\tiny \textbf{Cricket\_X}} 			  & 0.15 & 0.15 & \textbf{0.16} & \textbf{0.32} & 0.28 & 0.30 & \textbf{0.20} & 0.16 & 0.10 \\
\hline
{\tiny \textbf{ECG5000}} 				  & 0.29 & 0.28 & \textbf{0.31} & \textbf{0.32} & 0.29 & 0.30 & \textbf{0.24} & \textbf{0.24} & 0.16 \\
\hline
{\tiny \textbf{ElectricDevices}} 		  & 0.20 & \textbf{0.21} & 0.19 & \textbf{0.30} & 0.29 & 0.29 & \textbf{0.33} & 0.25 & 0.27 \\
\hline
{\tiny \textbf{FaceAll}} 				  & 0.15 & \textbf{0.21} & 0.15 & 0.33 & \textbf{0.36} & 0.29 & \textbf{0.22} & 0.15 & 0.09 \\
\hline
{\tiny \textbf{FordA}} 				  & \textbf{0.19} & 0.01 & 0.01 & \textbf{0.27} & 0.01 & 0.01 & \textbf{0.53} & 0.33 & 0.29 \\
\hline
{\tiny \textbf{InsectWingbeat}} 	  & 0.07 & \textbf{0.21} & 0.17 & 0.18 & \textbf{0.37} & 0.32 & \textbf{0.19} & 0.18 & 0.11 \\
\hline
{\tiny \textbf{UWaveGesture}}         & 0.17 & \textbf{0.2} & 0.18 & 0.27 & \textbf{0.28} & \textbf{0.28} & \textbf{0.2} & 0.19 & 0.12  \\
\hline
\hline
{\tiny \textbf{Synthetic (DGP)}}	  & 0.76 & 0.40 & \textbf{0.92} & 0.91 & 0.66 & \textbf{0.97} & \textbf{0.73} & 0.39 & 0.68 \\
\hline
{\tiny \textbf{Production Brazil}}	  & 0.09 & 0.18 & \textbf{0.30} & 0.40 & 0.55 & \textbf{0.70} & 0.46 & 0.39 & \textbf{0.61} \\ 

\hline
\end{tabular}
\label{table:4}
\end{table}

The results indicate that none of the three approaches performs  uniformly better than the others.  Some interesting comments follow. 
For the synthetic data sets  and 18Pairs,  \textit{tsfeatures} and \textit{NetF}  perform better than \textit{catch22}  in all  evaluation criteria. 
The clusters for ECG5000, ElectricDevices and UWaveGestureLibraryAll data sets produced by the three approaches fare equally well when  assessed  by ARI, NMI and AS. The same is true for M3 data and Cricket\_X data sets, with slightly lower results.
\textit{NetF} approach seems produce better clusters for CinC\_ECG\_torso measured according to the three criteria,  
the \textit{tsfeatures} seems to produce better clusters for FordA, 
and the \textit{catch22} for FaceAll and InsectWingbeat measured according to the ARI and NMI. 

Analyzing the overall results, Tables~\ref{table:4}, \ref{app:table_utsc_res1}-\ref{app:table_utsc_res3} we can state that \textit{tsfeatures} and \textit{NetF} approaches present the best  ARI and NMI evaluation metrics, while \textit{tsfeatures} achieves by far the best results in the AS.  
If we consider the UEA \& UCR repository classification of the data sets, we note the following: the  \textit{NetF} approach presents good results for time series data of the type Image (BeetleFly, FaceFour, MixedShapesRegularTrain, OSULeaf and Symbols),  ECG (CinC\_ECG\_tors and TwoLeadECG) and Sensor (Wafer); the \textit{tsfeatures} performs best for types Simulated (BME, UMD and TwoPatterns), ECG (NonInvasiveFetalECGTho), Image (ShapesAll) and Sensor (SonyAIBORobotSurface and Trace); finally  
 the \textit{catch22} approach presents best results for Spectro (Coffee), Device (HouseTwenty) and Simulated (ShapeletSim) types. In summary \textit{NetF} and \textit{tsfeatures} perform better in data with the same characteristics while \textit{catch22} seems to be more appropriate to capture other characteristics.

\begin{figure}[hbt!]
	\centering 
	\includegraphics[scale=.45]{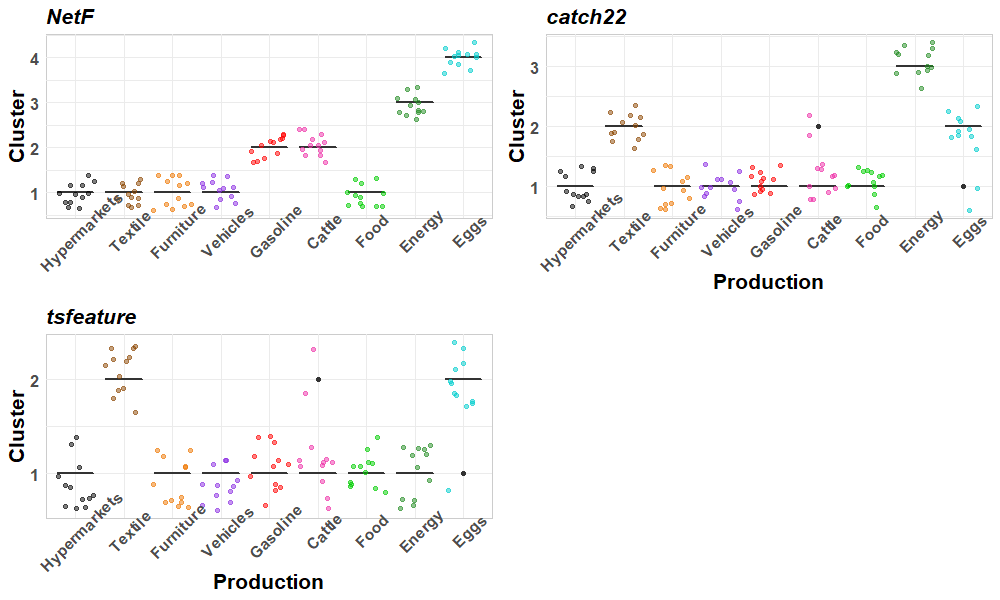}
	\caption{Attribution of the Production in Brazil time series to the different clusters, according to each of the feature approaches. 
	The different productions are represented on the horizontal axis and by a unique color. The time series are represented by the colored points according to its production type. The vertical axis represents the cluster number to which a time series is assigned.} 
	\label{Figure_7}
\end{figure}

\begin{figure}[hbt!]
	\centering 
	\includegraphics[scale=.43]{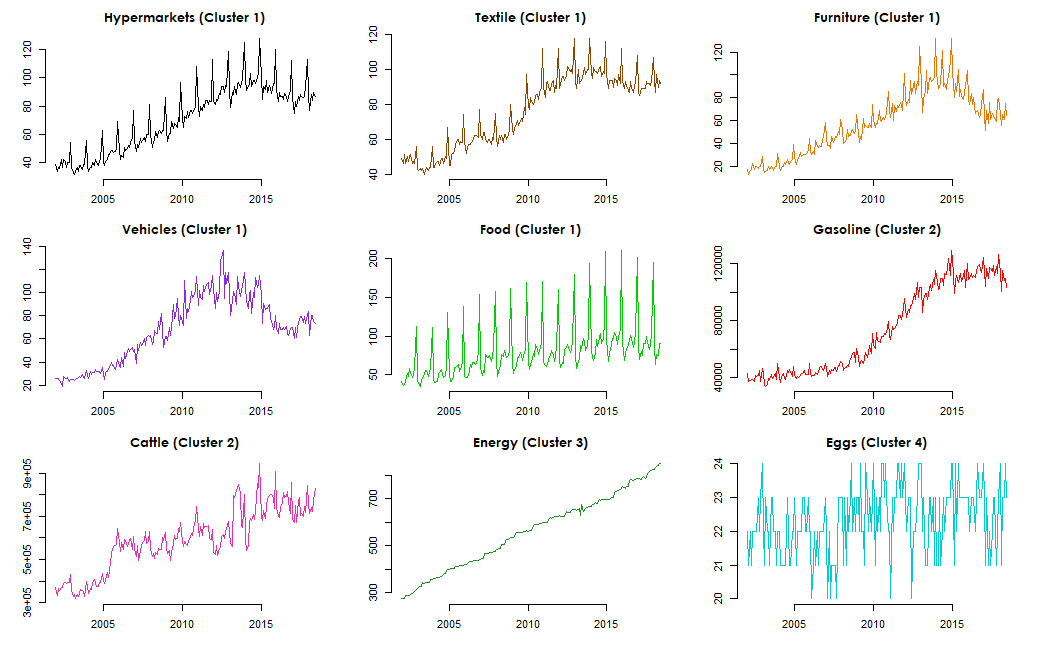}
	\caption{Production in Brazil representative of each cluster (indicated in subtitle) obtained using the proposed approach, \textit{NetF}.}
	\label{Figure_8}
\end{figure}

Regarding the data set Production in Brazil, Table~\ref{table:5} shows more detail on the clustering results, adding the value $k$ to indicate the number of clusters that was automatically computed. We note that the 4 clusters obtained with \textit{NetF} correspond to 4 types of goods: eggs; energy; gasoline and cattle; hypermarkets, textile, furniture, vehicles and food. Attribution plots of the clusters obtained by the three approaches are represented in Figure~\ref{Figure_7}. Note that both \textit{tsfeature} and \textit{catch22} put eggs and textile production in the same cluster, and \textit{tsfeature} cannot distinguish energy. 
Notice also how the \textit{NetF} approach produced the cluster with highest AS and hence the highest intra-cluster-similarity. To illustrate the relevance of the results, Figure~\ref{Figure_8} depicts a representative time series for each cluster.

\begin{table}[hbt!]
\centering
\caption{Clustering evaluation metrics for the different clustering analysis on Production in Brazil data set based on \textit{Netf}, \textit{tsfeatures} and \textit{catch22} approaches. 
The values reflect the mean of 10 repetitions of the proposed method for different feature vectors and for the number of clusters detected according to the average silhouette metric.}
\begin{tabular}{|l|c|c|c|c|}
\hline
\multicolumn{1}{|c|}{\multirow{2}{*}{\textbf{Approach}}} & \multicolumn{1}{c|}{\multirow{2}{*}{\textbf{$k$}}} & \multicolumn{1}{c|}{\textbf{ARI}} & \multicolumn{1}{c|}{\textbf{NMI}} & \multicolumn{1}{c|}{\textbf{AS}} \\
 & & \multicolumn{1}{c|}{\scriptsize $[-1,1]$} & \multicolumn{1}{c|}{\scriptsize $[0,1]$} & \multicolumn{1}{c|}{\scriptsize $[-1,1]$} \\
  
\hline
\textbf{\textit{NetF}}          	& 4 & \textbf{0.30} & \textbf{0.70} & \textbf{0.61} \\
\hline
\textbf{\textit{catch22}}        	& 3 & 0.18 & 0.55 & 0.39 \\
\hline
\textbf{\textit{tsfeatures}} 		& 2 & 0.09 & 0.40 & 0.46 \\
\hline

\end{tabular}
\label{table:5}
\end{table}

\newpage

\vspace{-0.4cm}

\newpage
\section{Conclusions}
\label{sec:concl}

In this paper we introduce \textit{NetF}, a novel set of 15 time series features, and we explore its ability to characterize time series data. Our methodology relies on mapping the time series into complex networks using three different mapping methods: natural and horizontal visibility and quantile graphs (based on transition probabilities). We then extract five topological measures for each mapped network, concatenating them into a single time series feature vector, and we describe in detail how we can do this in practice. 

To better understand the potential of our approach, we first perform an empirical evaluation on a synthetic data set of 3300 networks, grouped in 11 different and specific time series models. 
Analysing the weighted visibility (natural and horizontal) and quantile graphs feature space provided by \textit{NetF}, we were able to identify sets of features that distinguish non-stationary from stationary time series, counting from real-valued time series, periodic from non-periodic time series, state time series from non-state time series and heteroskedastic time series. 
 The non-stationarity time series have high values of average path length and low values of clustering coefficients in their QGs, and the opposite happens for the stationary time series. 
 Counting series have lowest value of average weighted  degree, highest value of number of communities in their QGs and lowest value of clustering coefficient in WHVGs, while the opposite happens for non-counting time series. 
For state time series the average weighted degree value in their weighted VGs is the lowest and the number of communities is high, the opposite happens for the non-state time series. 
Heteroskedastic time series are identified with high average weighted degree values of their WHVGs, compared to the other DGP's.

To further showcase the applicability of \textit{NetF}, we use its feature set for clustering both the previously mentioned synthetic data, as well as a large set of benchmark empirical time series data sets. 
The results for the data sets in which ground truth is available indicate that \textit{NetF} yields the highest mean for ARI (0.287) compared to alternative time series features, namely \textit{tsfeature} and \textit{catch22}, with  means of 0.267 and 0.228, respectively. 
For the NMI metric the results are similar (0.395, 0.397 and 0.358, respectively) and for AS the highest mean was found for \textit{tsfeature}, 0.332 versus approximately 0.3 for the others. 
However, the higher values for AS\footnote{ samples are very similar inside the cluster and show little similarity inter-cluster} must be viewed in light of the low values of ARI and NMI which indicate an imperfect formation of the clusters. 
For the production data in Brazil, for which no ground truth is available, \textit{NetF} produces  clusters which group production series with different characteristics, namely, time series of counts, marked upward trend, series in the same range of values, and with seasonal component. 

The results show that \textit{NetF} is capable of capturing information regarding specific properties of time series data. 
\textit{NetF} is also capable of grouping time series of different domains, such as data from ECGs, image and sensors, as well as identifying different characteristics of the time series using different mapping concepts, which stand out in different topological features.  
The general characteristics of the data, namely, the size of the data set, the length of the time series and the number of clusters, do not seem to be influencing the results obtained. 
Also, \textit{NetF} does not require typical time series preprocessing tasks, such as decomposing, differencing or whitening. 
Moreover, our methodology is applicable to any time series, regardless of the nature of the data. 

The mappings and topological network measures considered are global, but it is important to clarify that they do not constitute a  "universal" solution. 
In particular, we found that the weighted versions of the visibility graph mappings used here produce better results than their unweighted versions, as we can see in previous works~\cite{vanessa2018time}. 
In fact, formulating a set of general features capable of fully characterizing a time series  without knowing both the time series properties and the intended analysis is a difficult and challenging task~\cite{kang2020gratis}. 

For related future work, we intend to add and explore new sets of topological measures both, adding local and intermediate features to \textit{NetF}, as well as exploring other mapping methods (such as proximity graphs). We also intend to extend our approach to the multivariate case, since obtaining useful features for multidimensional time series analysis is still an open problem.

\section*{Availability of data and materials}

\textit{NetF} and data sets are available at \url{https://github.com/vanessa-silva/NetF}.
%\\Data sets are available at \url{https://www.dcc.fc.up.pt/~vanessa.silva/NetF}.

%%%%%%%%%%%%%%%%%%%%%%%%%%%%%%%%%%%%%%%%%%%%%%%%%%%%%%%%%%%%%%%%%%%

\begin{acknowledgements}
This work is financed by National Funds through the Portuguese funding agency, FCT - Funda\c{c}\~{a}o para a Ci\^{e}ncia e a Tecnologia, and UE - Uni\~{a}o Europeia, through national funds, and co-funded by the FSE - Fundo Social Europeu, MCTES - Minist\'{e}rio da Ci\^{e}ncia, Tecnologia e Ensino Superior, FEDER - Fundo Europeu de Desenvolvimento Regional within projects SFRH/BD/139630/2018 and LA/P/0063/2020.

The authors would like to thank Ana Paula Amazonas Soares of the Universidade Federal Rural de Pernambuco, Recife – Pernambuco, for providing the new Production in Brazil dataset.

The authors would like to thank the anonymous referees for their constructive comments which allowed to improve the paper.
\end{acknowledgements}

 \section*{Conflict of interest}
The authors declare that they have no conflict of interest.

\bibliographystyle{spmpsci}      % mathematics and physical sciences
\bibliography{bib}               % name your BibTeX data base

\newpage
\begin{appendices}
\renewcommand\thefigure{{\thesection.\arabic{figure}}}
\renewcommand\thetable{{\thesection.\arabic{table}}}

\section{Time Series Models}
\label{app:TSmodels}

\setcounter{figure}{0}
\setcounter{table}{0}

Main reference~\cite{Sumway2017}

\paragraph{\textbf{Linear Models}}.

\begin{description} 
    \item[\textbf{WN}] The \textit{white noise} process, $\epsilon_{t}$, is a sequence of i.i.d. random variables with mean 0 and constant variance $\sigma_{\epsilon}^{2}$. 
    It is the simplest time series process that reflects information that is neither directly observable nor predictable. 
    We generated $\epsilon_{t} \sim N(0,1)$ white noise processes and refer to them as \texttt{WN}. 
    \vspace{2.5mm}
    
    \item[\textbf{AR$(p)$}] We defined a process $Y_t$ as an AR process of order $p$ if it satisfies the following equation:
        \begin{equation} \label{eq_ar}
	        Y_{t} = \sum_{i=1}^{p} \phi_{i} Y_{t-i} + \epsilon_{t}, 
        \end{equation}
    where $\epsilon_{t}$ is the white noise and $\phi_{i}$ is the autoregressive constant. 
    We used $p \in \{1, 2\}$, and parameters $\phi_{1} \in \{ -0.5, 0.5\}$ to generate AR$(1)$ processes and $\phi_{1} = 1.5$ and $\phi_{2} = -0.75$ for AR$(2)$ processes. 
    These parameters ensure that the time series present the following characteristics: $\phi_{1} = 0.5$ leads to smoother time series than $\phi_{1} = -0.5$;  and $\phi_{1} = 1.5$ and $\phi_{2} = -0.75$ generates pseudo-periodic time series. 
	We refer to the three models generated as \texttt{AR(1)-0.5}, \texttt{AR(1)0.5} and \texttt{AR(2)}, respectively. 
    \vspace{2.5mm}
	
	\item[\textbf{ARIMA$(p,d,q)$}] The \textit{autoregressive integrated moving average} model is a generalization of the ARMA model suitable  
	for modeling non-stationary time series. 
	A process $Y_{t}$ is an ARMA$(p,q)$ process if it satisfies the equation:
	    \begin{equation} \label{eq_arima}
	        Y_{t} = \sum_{i=1}^{p} \phi_{i} Y_{t-i} + \sum_{i=1}^{q} \theta_{i} \epsilon_{t-i} + \epsilon_{t}, 
        \end{equation}
    where $\theta_{i}$ is the moving average constant. 
    If a process $Y_{t}$ is a non-stationary time series it can be written as an ARIMA$(p,d,q)$ process if its $d$th-differences $\nabla^{d} Y_{t} = (1 - B)^{d} Y_{t}$, $d \in \mathbb{N}$, is a stationary ARMA$(p,q)$ process. So $Y_{t}$ satisfies the following equation,
        \begin{equation} \label{eq_ar(f)ima}
            \left(1 - \sum_{i=1}^{p} \phi_{i} B^{i}\right) {\left(1-B\right)}^d Y_{t} = \left(1 + \sum_{i=1}^{q} \theta_{i} B^{i}\right) \epsilon_{t},
        \end{equation}
    where $B$ represents the backshift operator, $BY_{t} = Y_{t-1}$. 
	We use $p = 1$, $d = 1$, $q = 0$, and $\phi_{1} = 0.7$ to generate ARIMA$(1,1,0)$ processes with stochastic trend. 
	We refer to these processes as \texttt{ARIMA}.  
    \vspace{2.5mm}
	
    \item[\textbf{ARFIMA$(p,d,q)$}] \textit{Autoregressive fractionally integrated moving average} model is a generalization of the ARIMA model useful for modeling time series with long range dependence. 
	A process $Y_{t}$ is an ARFIMA$(p,d,q)$ process if it satisfies the (Eq~\ref{eq_ar(f)ima}) and the difference parameter, $d$, can take real values. 
    We generate ARFIMA$(1,0.4,0)$ processes that exhibit long memory and refer to them as \texttt{ARFIMA}. 
\end{description}

\newpage
Time series are generated from the above DGP using the R packages: \\ \texttt{timeSeries}~\cite{timeSeries} and \texttt{fracdiff}~\cite{fracdiff}.

\paragraph{\textbf{Non Linear Models}}

\begin{description}
    \item[\textbf{SETAR$(1)$}] The \textit{self-exciting threshold autoregressive} model of order $1$ specify the nonlinearity in the conditional mean. 
	It is useful for processes with regime changes that approximate a nonlinear function by piece wise linear functions dependent on the regime~\cite{Tong2011}. 
	This model can be presented by the following system of equations,
        \begin{equation} \label{eq_setar}
            Y_{t} = \left\{ \begin{array}{ll}
	           \alpha Y_{t-1} + \epsilon_{t} \ \ ,\ \ \mathrm{if}\ \ Y_{t-1} \leq r\\
	           \beta Y_{t-1} + \gamma\epsilon_{t} ,\ \ \mathrm{if}\ \  Y_{t-1} > r
            \end{array}\right.,
        \end{equation}
    where $r$ represents a real threshold. 
	We used  $\alpha = 0.5$, $\beta = -1.8$, $\gamma = 2$ and $r = -1$ and we generated time series with regimes with different autocorrelation properties: in the first the correlation is positive while in the second alternates between positive and negative values. 
	We refer to this model as \texttt{SETAR}. 
    \vspace{2.5mm}

    \item[\textbf{HMM}] \textit{Hidden Markov models} are probabilistic models for the joint probability the random variables $(Y_{1},\ldots,Y_{T},X_{1},\ldots,X_{T})$ where $Y_{t}$ is a discrete (or continuous) variable and $X_{t}$ is a \textit{hidden Markov chain} with a finite number of states, $N$. 
	The following conditional independence assumptions hold~\cite{Zucchini2016}:
	    \begin{enumerate}
	        \item $P(X_{t} \ | \ X_{t-1},Y_{t-1},\ldots,X_{1},Y_{1}) = P(X_t \ | \ X_{t-1})$,
	        \item $ P(Y_{t} \ | \ X_{T},Y_{T},\ldots,X_{1},Y_{1}) = P(Y_{t} \ | \ X_{t})$.
        \end{enumerate}
	We used $N=2$ and the transition matrix: 
	$\bigl[ \begin{smallmatrix}
  	0.9 & 0.1\\ 0.1 & 0.9
	\end{smallmatrix} \bigr]$. 
	The data are generated from a Poisson distribution with $\lambda = 10$ for the first regime and $\lambda = 15$ for the second. 
	We refer to this model as \texttt{HMM}. Time seriea are generated using the R package \texttt{HMMpa}~\cite{HMMpa}.
	
\end{description}

The next nonlinear models are based on ARCH models where the mean-corrected asset return is serially uncorrelated but dependent and  the dependency can be described by a simple quadratic function of its lagged values~\cite{Tsay2005}. 
Hereafter, $\epsilon_{t}$ are uncorrelated random variables, $z_{t}$ represents a white noise with variance 1 and $\sigma_{t}$ the standard deviation of $\epsilon_{t}$, that is $\epsilon_{t} = \sigma_{t} z_{t}$. 

\begin{description}
    \item[\textbf{GARCH$(p,q)$}] The GARCH model is a \textit{generalization} of the ARCH model in which the conditional volatility is a function not only of the squares of past innovations, but also of their own past values~\cite{Cryer2008}. 
	Thus, $\epsilon_{t}$ is a GARCH$(p,q)$ process if  it satisfies the following equation,
        \begin{equation} \label{eq_garch}
	        \sigma_{t}^{2} = \omega + \sum_{i=1}^{p} \beta_{i} \sigma_{t-i}^{2} + \sum_{i=1}^{q} \alpha_{i} \epsilon_{t-i}^{2},
        \end{equation}
    where $\omega > 0$, $\alpha_{i}, \beta_{i} \geq 0$, $\sum_{i=1}^{p} \beta_{i} + \sum_{i=1}^{q} \alpha_{i} < 1$. 
    The conditional standard deviation can exhibit persistent periods of high or low volatility because past values of the process are fed back into the present value. 
    We used $p = 1$, $q = 1$, $\omega = 10^{-6}$, $\alpha_{1} = 0.1$ and $\beta_{1} = 0.8$ to generate the GARCH$(1,1)$ processes and 
    we refer to them as \texttt{GARCH}. 
    \vspace{2.5mm}

\newpage    
    \item[\textbf{EGARCH$(p,q)$}]  
	The \textit{exponential} GARCH allows asymmetric effects of positive and negative shocks on volatility~\cite{Tsay2005}. 
	The EGARCH$(p,q)$ model is given by the equation,
        \begin{equation} \label{eq_egarch}
	        log(\sigma_t^{2}) = \omega + \sum_{i=1}^{q} \alpha_{i} \left|\frac{\epsilon_{t-i}}{\sigma_{t-i}}\right| + \sum_{i=1}^{p}\beta_{i} log(\sigma_{t-i}^{2}) + \sum_{i=1}^{q} \gamma_{i} \frac{\epsilon_{t-i}}{\sigma_{t-i}},
        \end{equation}
    where $\omega = \alpha_{0} - \alpha_{1} \sqrt{\frac{2}{\pi}}$,  $\alpha_{i}$ characterize the volatility clustering phenomena, $\beta_{i}$ is the persistence parameter, and $\gamma_{i}$ describes the leverage effect. 
    The logged conditional variance allows to relax the positivity constraint of the model coefficients. 
    To this model we choose $p = 1$, $q = 1$, $\omega = \left(10^{-6}-0.1 \sqrt{\frac{2}{\pi}}\right)$, $\alpha_{1} = 0.1$, $\beta_{1} = 0.01$ and $\gamma_{1} = 0.3$, and we refer to it as \texttt{EGARCH}. 
    \vspace{2.5mm}

    \item[\textbf{INAR$(1)$}] The \textit{integer-valued autoregressive} models have been proposed to model integer-valued time series, in particular, correlated counts~\cite{eduarda2004difference}. 
	These models are based on thinning (random) operations defined on the integers, where the following binomial thinning is the most common: let $X$ be a non-negative integer valued random variable and $0 < \alpha < 1$, then $\alpha \ast X = \sum_{i=1}^{X} Y_i$ where $\{Y_i\}$ is a sequence of i.i.d. Bernoulli random variables, independent of $X$. 
	A process $Y_{t}$ is said to be an INAR$(1)$ process if it satisfies the equation, 
        \begin{equation} \label{eq_inar}
	        Y_{t} = \alpha\ast Y_{t-1} + \epsilon_{t}.
        \end{equation}
    If the innovation sequence $\epsilon_t$ and the initial distribution are Poisson, $Y_t$ is said to be a Poisson INAR$(1)$ process. 
	We used $\alpha = 0.5$ and Poisson arrivals with $\epsilon_{t} \sim Po(1)$ to generate integer valued data with autocorrelation decaying at a rate of 0.5. 
	We refer to this model as \texttt{INAR}.
\end{description}

Time series from the \texttt{HMM} and \texttt{GARCH} are simulated using the R packages \texttt{HMMpa} \cite{HMMpa} and  \texttt{fGarch} \cite{fGarch}, respectively. 
Time series are generated from the remaining DGP of our own implementation, available from the authors.

\newpage
\section{ Feature evaluation in  Synthetic Time Series}
\label{app:netF_tsmodels}
\setcounter{figure}{0}
\setcounter{table}{0}

The topological features of WNVGs, WHVGs and QGs obtained from the 1100 time series models are, respectively, summarized in tables~\ref{app_table:1}, \ref{app_table:2} and \ref{app_table:3}. 
The table reporting the mean and standard deviation (in brackets) of the \textit{Min-Max} normalized (across models) metrics. The columns of the tables are colored with a gradient based on the mean values: cells with a maximum value 1 are colored red, cells with the minimum value 0 are colored white and the remainder with a hue of red color proportional to its value. 

\subsection*{\textbf{\textit{Weighted Natural Visibility Graphs}}}
\vspace*{-0.25in}
\begin{table}[!ht]
\caption[Topological metrics of WNVGs from time series models]{Table of mean values of the $100$ instances of each DGP for each topological metric, resulting from WNVGs. The  standard deviations are presented in parentheses.}
\label{app_table:1}
\centering

\begin{tabular}{|>{\bfseries\leavevmode\color{black}}l|>{\leavevmode\color{black}}c|>{\leavevmode\color{black}}c|>{\leavevmode\color{black}}c|>{\leavevmode\color{black}}c|>{\leavevmode\color{black}}c|}

\hline
\multicolumn{1}{|c|}{\textcolor{black}{\multirow{3}{*}{\textbf{\small{Models}}}}} & \multicolumn{1}{|c|}{\textcolor{black}{\textbf{\scriptsize{Average}}}} & \multicolumn{1}{|c|}{\textcolor{black}{\textbf{\scriptsize{Average}}}} & \multicolumn{1}{|c|}{\textcolor{black}{\textbf{\scriptsize{Number of}}}} & \multicolumn{1}{|c|}{\textcolor{black}{\textbf{\scriptsize{Clustering}}}} & \multicolumn{1}{|c|}{\textcolor{black}{\multirow{2}{*}{\textbf{\scriptsize{Modularity}}}}}\\

\multicolumn{1}{|c|}{\textcolor{black}{\textbf{ }}} & \multicolumn{1}{|c|}{\textcolor{black}{\textbf{\scriptsize{Degree}}}} & \multicolumn{1}{|c|}{\textcolor{black}{\textbf{\scriptsize{Path Length}}}} & \multicolumn{1}{|c|}{\textcolor{black}{\textbf{\scriptsize{Communities}}}} & \multicolumn{1}{|c|}{\textcolor{black}{\textbf{\scriptsize{Coefficient}}}} & \multicolumn{1}{|c|}{\textcolor{black}{\textbf{}}}\\

\multicolumn{1}{|c|}{\textcolor{black}{\textbf{ }}} & \multicolumn{1}{|c|}{\textcolor{black}{\textbf{\scriptsize{($\bar{k}$)}}}} & \multicolumn{1}{|c|}{\textcolor{black}{\textbf{\scriptsize{($\bar{d}$)}}}} & \multicolumn{1}{|c|}{\textcolor{black}{\textbf{\scriptsize{($S$)}}}} & \multicolumn{1}{|c|}{\textcolor{black}{\textbf{\scriptsize{($C$)}}}} & \multicolumn{1}{|c|}{\textcolor{black}{\textbf{($Q$)}}}\\

\hline
\multirow{2}{*}{{\texttt{AR(1)-0.5}}}
& \cellcolor{ff9292}0.430 & \cellcolor{ff0000}0.457  & \cellcolor{ffaeae}0.225 & \cellcolor{ff7373}0.615 & \cellcolor{ff2525}0.900 \\

& \cellcolor{ff9292}\scriptsize{(0.006)} & \cellcolor{ff0000}\scriptsize{(0.032)} & \cellcolor{ffaeae}\scriptsize{(0.063)} & \cellcolor{ff7373}\scriptsize{(0.007)} & \cellcolor{ff2525}\scriptsize{(0.022)}\\

\hline
\multirow{2}{*}{{\texttt{AR(1)0.5}}}
& \cellcolor{ff4c4c}0.698 & \cellcolor{ff0e0e}0.443 & \cellcolor{ffe5e5}0.116 & \cellcolor{ff4646}0.751 & \cellcolor{ff0b0b}0.963 \\

& \cellcolor{ff4c4c}\scriptsize{(0.005)} & \cellcolor{ff0e0e}\scriptsize{(0.037)} & \cellcolor{ffe5e5}\scriptsize{(0.037)} & \cellcolor{ff4646}\scriptsize{(0.009)} & \cellcolor{ff0b0b}\scriptsize{(0.010)}\\

\hline
\multirow{2}{*}{{\texttt{AR(2)}}}
& \cellcolor{ff4747}0.719 & \cellcolor{ff3232}0.408 & \cellcolor{ff3131}0.472 & \cellcolor{ff0000}0.968 & \cellcolor{ff5151}0.792 \\

& \cellcolor{ff4747}\scriptsize{(0.008)} & \cellcolor{ff3232}\scriptsize{(0.035)} & \cellcolor{ff3131}\scriptsize{(0.132)} & \cellcolor{ff0000}\scriptsize{(0.012)} & \cellcolor{ff5151}\scriptsize{(0.082)}\\

\hline
\multirow{2}{*}{{\texttt{ARIMA}}}
& \cellcolor{ff1313}0.919 & \cellcolor{ffffff}0.211 & \cellcolor{ff9e9e}0.257 & \cellcolor{ffffff}0.188 & \cellcolor{ffffff}0.367 \\

& \cellcolor{ff1313}\scriptsize{(0.006)} & \cellcolor{ffffff}\scriptsize{(0.095)} & \cellcolor{ff9e9e}\scriptsize{(0.115)} & \cellcolor{ffffff}\scriptsize{(0.079)} & \cellcolor{ffffff}\scriptsize{(0.140)}\\

\hline
\multirow{2}{*}{{\texttt{ARFIMA}}}
& \cellcolor{ff3434}0.790 & \cellcolor{ff2e2e}0.412 & \cellcolor{ffeeee}0.099 & \cellcolor{ff4242}0.766 & \cellcolor{ff0707}0.973 \\

& \cellcolor{ff3434}\scriptsize{(0.005)} & \cellcolor{ff2e2e}\scriptsize{(0.036)} & \cellcolor{ffeeee}\scriptsize{(0.035)} & \cellcolor{ff4242}\scriptsize{(0.012)} & \cellcolor{ff0707}\scriptsize{(0.009)}\\

\hline
\multirow{2}{*}{{\texttt{SETAR}}}
& \cellcolor{ffbbbb}0.273 & \cellcolor{ff1313}0.438  & \cellcolor{ff2727}0.491 & \cellcolor{ff6e6e}0.631 & \cellcolor{ff5959}0.772 \\

& \cellcolor{ffbbbb}\scriptsize{(0.006)} & \cellcolor{ff1313}\scriptsize{(0.041)} & \cellcolor{ff2727}\scriptsize{(0.145)} & \cellcolor{ff6e6e}\scriptsize{(0.007)} & \cellcolor{ff5959}\scriptsize{(0.070)}\\

\hline
\multirow{2}{*}{{\texttt{HMM}}}
& \cellcolor{ffffff}0.012 & \cellcolor{ff0b0b}0.446 & \cellcolor{ff0000}0.570 & \cellcolor{ff6262}0.667 & \cellcolor{ff8989}0.654 \\

& \cellcolor{ffffff}\scriptsize{(0.005)} & \cellcolor{ff0b0b}\scriptsize{(0.039)} & \cellcolor{ff0000}\scriptsize{(0.150)} & \cellcolor{ff6262}\scriptsize{(0.008)} & \cellcolor{ff8989}\scriptsize{(0.093)}\\

\hline
\multirow{2}{*}{{\texttt{INAR}}}
& \cellcolor{ff6e6e}0.570 & \cellcolor{ff0505}0.452 & \cellcolor{ffdede}0.131 & \cellcolor{ff6e6e}0.631 & \cellcolor{ff0e0e}0.956 \\

& \cellcolor{ff6e6e}\scriptsize{(0.004)} & \cellcolor{ff0505}\scriptsize{(0.118)} & \cellcolor{ffdede}\scriptsize{(0.052)} & \cellcolor{ff6e6e}\scriptsize{(0.011)}& \cellcolor{ff0e0e}\scriptsize{(0.024)}\\

\hline
\multirow{2}{*}{{\texttt{GARCH}}}
& \cellcolor{ff0000}0.994 & \cellcolor{ff1818}0.433 & \cellcolor{fffcfc}0.070 & \cellcolor{ff6767}0.652 & \cellcolor{ff0000}0.991 \\

& \cellcolor{ff0000}\scriptsize{(0.003)} & \cellcolor{ff1818}\scriptsize{(0.036)} & \cellcolor{fffcfc}\scriptsize{(0.022)} & \cellcolor{ff6767}\scriptsize{(0.010)} & \cellcolor{ff0000}\scriptsize{(0.005)}\\

\hline
\multirow{2}{*}{{\texttt{EGARCH}}}
& \cellcolor{ff0e0e}0.940 & \cellcolor{ff2121}0.425 & \cellcolor{ffffff}0.066 & \cellcolor{ff5f5f}0.677 & \cellcolor{ff0404}0.980 \\

& \cellcolor{ff0e0e}\scriptsize{(0.004)} & \cellcolor{ff2121}\scriptsize{(0.032)} & \cellcolor{ffffff}\scriptsize{(0.027)} & \cellcolor{ff5f5f}\scriptsize{(0.007)} & \cellcolor{ff0404}\scriptsize{(0.004)}\\

\hline
\multirow{2}{*}{{\texttt{WN}}}
& \cellcolor{ff6b6b}0.581 & \cellcolor{ff0707}0.450 & \cellcolor{ffdfdf}0.128 & \cellcolor{ff6262}0.667 & \cellcolor{ff1414}0.942 \\

& \cellcolor{ff6b6b}\scriptsize{(0.005)} & \cellcolor{ff0707}\scriptsize{(0.034)} & \cellcolor{ffdfdf}\scriptsize{(0.044)} & \cellcolor{ff6262}\scriptsize{(0.007)} & \cellcolor{ff1414}\scriptsize{(0.009)}\\

\hline
\end{tabular}

\end{table}

\newpage
\subsection*{\textbf{\textit{Weighted Horizontal Visibility Graphs}}}
\vspace*{-0.25in}

\begin{table}[!ht]
\caption[Topological metrics of WHVGs from time series models]{Table of mean values of the $100$ instances of each DGP for each topological metric, resulting from WHVGs. The  standard deviations are presented in parentheses.}
\label{app_table:2}
\centering

\begin{tabular}{|>{\bfseries\leavevmode\color{black}}l|>{\leavevmode\color{black}}c|>{\leavevmode\color{black}}c|>{\leavevmode\color{black}}c|>{\leavevmode\color{black}}c|>{\leavevmode\color{black}}c|}

\hline
\multicolumn{1}{|c|}{\textcolor{black}{\multirow{3}{*}{\textbf{\small{Models}}}}} & \multicolumn{1}{|c|}{\textcolor{black}{\textbf{\scriptsize{Average}}}} & \multicolumn{1}{|c|}{\textcolor{black}{\textbf{\scriptsize{Average}}}} & \multicolumn{1}{|c|}{\textcolor{black}{\textbf{\scriptsize{Number of}}}} & \multicolumn{1}{|c|}{\textcolor{black}{\textbf{\scriptsize{Clustering}}}} & \multicolumn{1}{|c|}{\textcolor{black}{\multirow{2}{*}{\textbf{\scriptsize{Modularity}}}}}\\

\multicolumn{1}{|c|}{\textcolor{black}{\textbf{ }}} & \multicolumn{1}{|c|}{\textcolor{black}{\textbf{\scriptsize{Degree}}}} & \multicolumn{1}{|c|}{\textcolor{black}{\textbf{\scriptsize{Path Length}}}} & \multicolumn{1}{|c|}{\textcolor{black}{\textbf{\scriptsize{Communities}}}} & \multicolumn{1}{|c|}{\textcolor{black}{\textbf{\scriptsize{Coefficient}}}} & \multicolumn{1}{|c|}{\textcolor{black}{\textbf{}}}\\

\multicolumn{1}{|c|}{\textcolor{black}{\textbf{ }}} & \multicolumn{1}{|c|}{\textcolor{black}{\textbf{\scriptsize{($\bar{k}$)}}}} & \multicolumn{1}{|c|}{\textcolor{black}{\textbf{\scriptsize{($\bar{d}$)}}}} & \multicolumn{1}{|c|}{\textcolor{black}{\textbf{\scriptsize{($S$)}}}} & \multicolumn{1}{|c|}{\textcolor{black}{\textbf{\scriptsize{($C$)}}}} & \multicolumn{1}{|c|}{\textcolor{black}{\textbf{($Q$)}}}\\

\hline
\multirow{2}{*}{{\texttt{AR(1)-0.5}}}
& \cellcolor{ff8787}0.473  & \cellcolor{fffefe}0.004  & \cellcolor{ffc3c3}0.230  & \cellcolor{ff7373}0.557  & \cellcolor{ffe0e0}0.319 \\

& \cellcolor{ff8787}\scriptsize{(0.005)} & \cellcolor{fffefe}\scriptsize{(0.001)} & \cellcolor{ffc3c3}\scriptsize{(0.057)} & \cellcolor{ff7373}\scriptsize{(0.004)} & \cellcolor{ffe0e0}\scriptsize{(0.056)}\\

\hline
\multirow{2}{*}{{\texttt{AR(1)0.5}}}
& \cellcolor{ff5f5f}0.628  & \cellcolor{fffcfc}0.009  & \cellcolor{ffd7d7}0.174  & \cellcolor{ff4d4d}0.697  & \cellcolor{ff3131}0.793 \\

& \cellcolor{ff5f5f}\scriptsize{(0.003)} & \cellcolor{fffcfc}\scriptsize{(0.001)} & \cellcolor{ffd7d7}\scriptsize{(0.046)} & \cellcolor{ff4d4d}\scriptsize{(0.004)} & \cellcolor{ff3131}\scriptsize{(0.031)}\\

\hline
\multirow{2}{*}{{\texttt{AR(2)}}}
& \cellcolor{ff9090}0.438  & \cellcolor{fff6f6}0.022  & \cellcolor{ff4343}0.600  & \cellcolor{ff0909}0.953  & \cellcolor{ffc3c3}0.397 \\

& \cellcolor{ff9090}\scriptsize{(0.006)} & \cellcolor{fff6f6}\scriptsize{(0.002)} & \cellcolor{ff4343}\scriptsize{(0.092)} & \cellcolor{ff0909}\scriptsize{(0.004)} & \cellcolor{ffc3c3}\scriptsize{(0.074)}\\

\hline
\multirow{2}{*}{{\texttt{ARIMA}}}
& \cellcolor{ff9696}0.414  & \cellcolor{ff0000}0.588  & \cellcolor{ff0000}0.794  & \cellcolor{ff0000}0.988  & \cellcolor{ffffff}0.236 \\

& \cellcolor{ff9696}\scriptsize{(0.007)} & \cellcolor{ff0000}\scriptsize{(0.161)} & \cellcolor{ff0000}\scriptsize{(0.106)} & \cellcolor{ff0000}\scriptsize{(0.005)} & \cellcolor{ffffff}\scriptsize{(0.074)}\\

\hline
\multirow{2}{*}{{\texttt{ARFIMA}}}
& \cellcolor{ff5e5e}0.633  & \cellcolor{fff3f3}0.030  & \cellcolor{ffcccc}0.204  & \cellcolor{ff3636}0.785  & \cellcolor{ff1212}0.878 \\

& \cellcolor{ff5e5e}\scriptsize{(0.003)} & \cellcolor{fff3f3}\scriptsize{(0.004)} & \cellcolor{ffcccc}\scriptsize{(0.047)} & \cellcolor{ff3636}\scriptsize{(0.004)} & \cellcolor{ff1212}\scriptsize{(0.028)}\\

\hline
\multirow{2}{*}{{\texttt{SETAR}}}
& \cellcolor{ffb8b8}0.284  & \cellcolor{ffffff}0.003   & \cellcolor{ff6363}0.508  & \cellcolor{ff8181}0.504  & \cellcolor{ffd3d3}0.354 \\

& \cellcolor{ffb8b8}\scriptsize{(0.005)} & \cellcolor{ffffff}\scriptsize{(0.001)} & \cellcolor{ff6363}\scriptsize{(0.086)} & \cellcolor{ff8181}\scriptsize{(0.004)} & \cellcolor{ffd3d3}\scriptsize{(0.092)}\\

\hline
\multirow{2}{*}{{\texttt{HMM}}}
& \cellcolor{ffffff}0.009  & \cellcolor{fffafa}0.013  & \cellcolor{ff3535}0.640  & \cellcolor{ff8b8b}0.467  & \cellcolor{ffc8c8}0.385 \\

& \cellcolor{ffffff}\scriptsize{(0.003)} & \cellcolor{fffafa}\scriptsize{(0.002)} & \cellcolor{ff3535}\scriptsize{(0.116)} & \cellcolor{ff8b8b}\scriptsize{(0.006)} & \cellcolor{ffc8c8}\scriptsize{(0.105)}\\

\hline
\multirow{2}{*}{{\texttt{INAR}}}
& \cellcolor{ff9999}0.403  & \cellcolor{fff1f1}0.034  & \cellcolor{ffaeae}0.292  & \cellcolor{ffffff}0.034  & \cellcolor{ff0000}0.927 \\

& \cellcolor{ff9999}\scriptsize{(0.002)} & \cellcolor{fff1f1}\scriptsize{(0.005)} & \cellcolor{ffaeae}\scriptsize{(0.056)} & \cellcolor{ffffff}\scriptsize{(0.010)}& \cellcolor{ff0000}\scriptsize{(0.033)}\\

\hline
\multirow{2}{*}{{\texttt{GARCH}}}
& \cellcolor{ff0000}0.998  & \cellcolor{fffdfd}0.007  & \cellcolor{fffdfd}0.064  & \cellcolor{ff6464}0.611  & \cellcolor{ff3737}0.777 \\

& \cellcolor{ff0000}\scriptsize{(0.001)} & \cellcolor{fffdfd}\scriptsize{(0.002)} & \cellcolor{fffdfd}\scriptsize{(0.022)} & \cellcolor{ff6464}\scriptsize{(0.004)} & \cellcolor{ff3737}\scriptsize{(0.015)}\\

\hline
\multirow{2}{*}{{\texttt{EGARCH}}}
& \cellcolor{ff1919}0.900 & \cellcolor{fffefe}0.004 & \cellcolor{ffffff}0.059 & \cellcolor{ff6a6a}0.591 & \cellcolor{ff4242}0.747\\

& \cellcolor{ff1919}\scriptsize{(0.001)}  & \cellcolor{fffefe}\scriptsize{(0.001)}  & \cellcolor{ffffff}\scriptsize{(0.026)}  & \cellcolor{ff6a6a}\scriptsize{(0.005)}  & \cellcolor{ff4242}\scriptsize{(0.017)} \\

\hline
\multirow{2}{*}{{\texttt{WN}}}
& \cellcolor{ff6a6a}0.584 & \cellcolor{fffefe}0.004  & \cellcolor{ffd8d8}0.170  & \cellcolor{ff6666}0.605  & \cellcolor{ff6f6f}0.624 \\

& \cellcolor{ff6a6a}\scriptsize{(0.004)} & \cellcolor{fffefe}\scriptsize{(0.001)} & \cellcolor{ffd8d8}\scriptsize{(0.045)} & \cellcolor{ff6666}\scriptsize{(0.004)} & \cellcolor{ff6f6f}\scriptsize{(0.033)}\\

\hline
\end{tabular}

\end{table}

\newpage
\subsection*{\textbf{\textit{Quantile Graphs}}}
\vspace*{-0.25in}

\begin{table}[!ht]
\caption[Topological metrics of 50-QGs from time series models]{Table of mean values of the $100$ instances of each DGP for each topological metric, resulting from QGs. The  standard deviations are presented in parentheses.}
\label{app_table:3}
\centering

\begin{tabular}{|>{\bfseries\leavevmode\color{black}}l|>{\leavevmode\color{black}}c|>{\leavevmode\color{black}}c|>{\leavevmode\color{black}}c|>{\leavevmode\color{black}}c|>{\leavevmode\color{black}}c|}

\hline
\multicolumn{1}{|c|}{\textcolor{black}{\multirow{3}{*}{\textbf{\small{Models}}}}} & \multicolumn{1}{|c|}{\textcolor{black}{\textbf{\scriptsize{Average}}}} & \multicolumn{1}{|c|}{\textcolor{black}{\textbf{\scriptsize{Average}}}} & \multicolumn{1}{|c|}{\textcolor{black}{\textbf{\scriptsize{Number of}}}} & \multicolumn{1}{|c|}{\textcolor{black}{\textbf{\scriptsize{Clustering}}}} & \multicolumn{1}{|c|}{\textcolor{black}{\multirow{2}{*}{\textbf{\scriptsize{Modularity}}}}}\\

\multicolumn{1}{|c|}{\textcolor{black}{\textbf{ }}} & \multicolumn{1}{|c|}{\textcolor{black}{\textbf{\scriptsize{Degree}}}} & \multicolumn{1}{|c|}{\textcolor{black}{\textbf{\scriptsize{Path Length}}}} & \multicolumn{1}{|c|}{\textcolor{black}{\textbf{\scriptsize{Communities}}}} & \multicolumn{1}{|c|}{\textcolor{black}{\textbf{\scriptsize{Coefficient}}}} & \multicolumn{1}{|c|}{\textcolor{black}{\textbf{}}}\\

\multicolumn{1}{|c|}{\textcolor{black}{\textbf{ }}} & \multicolumn{1}{|c|}{\textcolor{black}{\textbf{\scriptsize{($\bar{k}$)}}}} & \multicolumn{1}{|c|}{\textcolor{black}{\textbf{\scriptsize{($\bar{d}$)}}}} & \multicolumn{1}{|c|}{\textcolor{black}{\textbf{\scriptsize{($S$)}}}} & \multicolumn{1}{|c|}{\textcolor{black}{\textbf{\scriptsize{($C$)}}}} & \multicolumn{1}{|c|}{\textcolor{black}{\textbf{($Q$)}}}\\

\hline
\multirow{2}{*}{{\texttt{AR(1)-0.5}}}
& \cellcolor{ff0000}1.000   & \cellcolor{fffdfd}0.005   & \cellcolor{ffffff}0.000   & \cellcolor{ff0808}0.972   & \cellcolor{ffffff}0.008 \\

& \cellcolor{ff0000}\scriptsize{(0.000)} & \cellcolor{fffdfd}\scriptsize{(0.000)} & \cellcolor{ffffff}\scriptsize{(0.000)} & \cellcolor{ff0808}\scriptsize{(0.003)} & \cellcolor{ffffff}\scriptsize{(0.003)}\\

\hline
\multirow{2}{*}{{\texttt{AR(1)0.5}}}
& \cellcolor{ff0000}1.000   & \cellcolor{fffdfd}0.005   & \cellcolor{fff7f7}0.027   & \cellcolor{ff0808}0.971   & \cellcolor{ff9595}0.374 \\

& \cellcolor{ff0000}\scriptsize{(0.000)} & \cellcolor{fffdfd}\scriptsize{(0.000)} & \cellcolor{fff7f7}\scriptsize{(0.010)} & \cellcolor{ff0808}\scriptsize{(0.003)} & \cellcolor{ff9595}\scriptsize{(0.044)}\\

\hline
\multirow{2}{*}{{\texttt{AR(2)}}}
& \cellcolor{ff0000}1.000   & \cellcolor{fff8f8}0.024   & \cellcolor{fff3f3}0.044   & \cellcolor{ff3232}0.829   & \cellcolor{ff1515}0.816 \\

& \cellcolor{ff0000}\scriptsize{(0.000)} & \cellcolor{fff8f8}\scriptsize{(0.000)} & \cellcolor{fff3f3}\scriptsize{(0.015)} & \cellcolor{ff3232}\scriptsize{(0.003)} & \cellcolor{ff1515}\scriptsize{(0.028)}\\

\hline
\multirow{2}{*}{{\texttt{ARIMA}}}
& \cellcolor{ff0000}1.000   & \cellcolor{ff0000}0.943   & \cellcolor{ffeeee}0.062   & \cellcolor{ffffff}0.144   & \cellcolor{ff8e8e}0.398 \\

& \cellcolor{ff0000}\scriptsize{(0.000)} & \cellcolor{ff0000}\scriptsize{(0.068)} & \cellcolor{ffeeee}\scriptsize{(0.022)} & \cellcolor{ffffff}\scriptsize{(0.132)} & \cellcolor{ff8e8e}\scriptsize{(0.099)}\\

\hline
\multirow{2}{*}{{\texttt{ARFIMA}}}
& \cellcolor{ff0000}1.000   & \cellcolor{fff7f7}0.029   & \cellcolor{fff2f2}0.047   & \cellcolor{ff3939}0.808   & \cellcolor{ff0000}0.890 \\

& \cellcolor{ff0000}\scriptsize{(0.000)} & \cellcolor{fff7f7}\scriptsize{(0.003)} & \cellcolor{fff2f2}\scriptsize{(0.017)} & \cellcolor{ff3939}\scriptsize{(0.009)} & \cellcolor{ff0000}\scriptsize{(0.039)}\\

\hline
\multirow{2}{*}{{\texttt{SETAR}}}
& \cellcolor{ff0000}1.000   & \cellcolor{fffafa}0.016    & \cellcolor{fff7f7}0.027   & \cellcolor{ff1010}0.946   & \cellcolor{ffbaba}0.245 \\

& \cellcolor{ff0000}\scriptsize{(0.000)} & \cellcolor{fffafa}\scriptsize{(0.000)} & \cellcolor{fff7f7}\scriptsize{(0.009)} & \cellcolor{ff1010}\scriptsize{(0.003)} & \cellcolor{ffbaba}\scriptsize{(0.039)}\\

\hline
\multirow{2}{*}{{\texttt{HMM}}}
& \cellcolor{ffb8b8}0.276   & \cellcolor{ffffff}0.001   & \cellcolor{ff4141}0.730   & \cellcolor{ff0000}0.998   & \cellcolor{ffbbbb}0.289 \\

& \cellcolor{ffb8b8}\scriptsize{(0.008)} & \cellcolor{ffffff}\scriptsize{(0.000)} & \cellcolor{ff4141}\scriptsize{(0.008)} & \cellcolor{ff0000}\scriptsize{(0.003)} & \cellcolor{ffbbbb}\scriptsize{(0.027)}\\

\hline
\multirow{2}{*}{{\texttt{INAR}}}
& \cellcolor{ffffff}0.000  & \cellcolor{fffefe}0.002   & \cellcolor{ff0000}0.981   & \cellcolor{ff0404}0.984   & \cellcolor{ff6565}0.493 \\

& \cellcolor{ffffff}\scriptsize{(0.002)} & \cellcolor{fffefe}\scriptsize{(0.001)} & \cellcolor{ff0000}\scriptsize{(0.009)} & \cellcolor{ff0404}\scriptsize{(0.024)}& \cellcolor{ff6565}\scriptsize{(0.010)}\\

\hline
\multirow{2}{*}{{\texttt{GARCH}}}
& \cellcolor{ff0000}1.000   & \cellcolor{ffffff}0.001   & \cellcolor{fff6f6}0.031   & \cellcolor{ff0000}1.000   & \cellcolor{fff1f1}0.055 \\

& \cellcolor{ff0000}\scriptsize{(0.000)} & \cellcolor{ffffff}\scriptsize{(0.000)} & \cellcolor{fff6f6}\scriptsize{(0.012)} & \cellcolor{ff0000}\scriptsize{(0.001)} & \cellcolor{fff1f1}\scriptsize{(0.016)}\\

\hline
\multirow{2}{*}{{\texttt{EGARCH}}}
& \cellcolor{ff0000}1.000 & \cellcolor{fffefe}0.002 & \cellcolor{fffafa}0.019 & \cellcolor{ff0000}0.999 & \cellcolor{fff5f5}0.041 \\

& \cellcolor{ff0000}\scriptsize{(0.000)} & \cellcolor{fffefe}\scriptsize{(0.000)} & \cellcolor{fffafa}\scriptsize{(0.010)} & \cellcolor{ff0000}\scriptsize{(0.001)} & \cellcolor{fff5f5}\scriptsize{(0.016)} \\

\hline
\multirow{2}{*}{{\texttt{WN}}}
& \cellcolor{ff0000}1.000  & \cellcolor{ffffff}0.001   & \cellcolor{fff7f7}0.029   & \cellcolor{ff0000}1.000   & \cellcolor{fff3f3}0.047 \\

& \cellcolor{ff0000}\scriptsize{(0.000)} & \cellcolor{ffffff}\scriptsize{(0.000)} & \cellcolor{fff7f7}\scriptsize{(0.011)} & \cellcolor{ff0000}\scriptsize{(0.000)} & \cellcolor{fff3f3}\scriptsize{(0.011)}\\

\hline
\end{tabular}

\end{table}

\subsection*{\textbf{\textit{Principal Component Analysis Results}}}
\vspace*{-0.25in}

\begin{figure}[hbt!]
	\centering 
	\includegraphics[scale=.37]{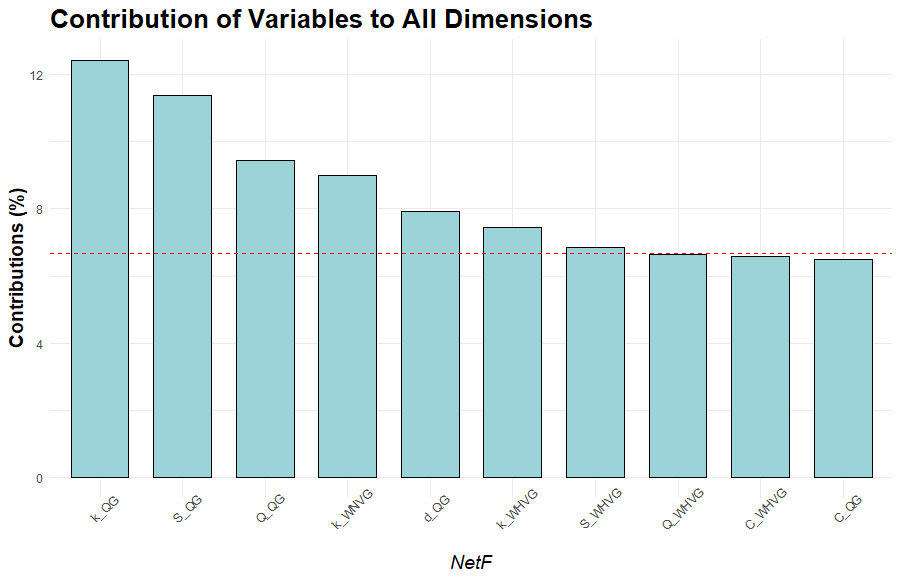}
	\caption{Bar plot with contributions of \textit{NetF} features to the total of all 15 principal components formed by the PCA. 
	The red dashed line on the plot indicates the expected average contribution. }
	\label{app_fig2}
\end{figure}

\newpage
\section{Clustering of Time Series Models}
\label{app:tsmodels_clust}

\begin{figure}[hbt!]
	\centering 
	\includegraphics[scale=.4]{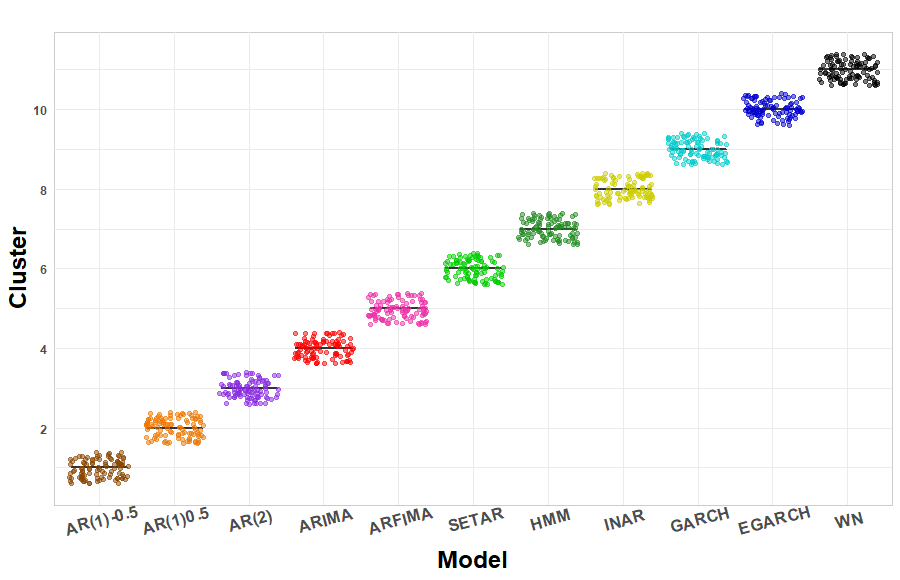}
	\caption{Attribution of the samples corresponding to instances of time series models to the different clusters, according to \textit{NetF}. 
	The different models are represented on the horizontal axis and by a unique color. The time series are represented by the colored points according to its model process. The vertical axis represents the cluster number to which a time series is assigned.}
	\label{Figure_4}
\end{figure}

We analyse the performance of different combinations of the feature vectors from the WNVG, WHVG and QG mappings in a clustering task using the synthetic data set. 
We set the  number of clusters to $k=11,$ the total of time series models, and assess the clustering results with the evaluation metrics. 
The results summarized in Table~\ref{table:3}\footnote{The results are means from 10 repetitions of the clustering analysis. The corresponding standard deviations indicate little or none variation between the repetitions.}
 indicate  that joining the features obtained from the two mapping concepts (VGs and QGs) 
 adds information that leads to
 improvements in the clustering results (compare the first three rows of the Table~\ref{table:3} with the last three). 
 In fact, as illustrated in Figure~\ref{Figure_4}, clustering based on \textit{NetF} can leads to a perfect attribution of the time series models samples across the 11 different clusters.
 
\begin{table}[hbt!]
\centering
\caption{Clustering evaluation metrics for the different clustering analysis resulting from different network-based feature vectors. 
The values reflect the mean of 10 repetitions of the proposed method for different feature vectors and for the ground truth ($k = 11$). The highest values are highlighted.}
\begin{tabular}{|l|r|r|r|}
\hline
\multicolumn{1}{|c|}{\multirow{2}{*}{\textbf{Mappings}}} & \multicolumn{1}{c|}{\textbf{ARI}} & \multicolumn{1}{c|}{\textbf{NMI}} & \multicolumn{1}{c|}{\textbf{AS}} \\
 & \multicolumn{1}{c|}{\scriptsize $[-1,1]$} & \multicolumn{1}{c|}{\scriptsize $[0,1]$} & \multicolumn{1}{c|}{\scriptsize $[-1,1]$} \\
  
\hline
\textbf{WNVG} 					& 0.68 & 0.86 & 0.51 \\
\hline
\textbf{WHVG} 					& 0.83 & 0.94 & 0.63 \\
\hline
\textbf{QG} 					& 0.64 & 0.84 & 0.66 \\
\hline
\textbf{WNVG - WHVG}    		& 0.81 & 0.93 & 0.57 \\
\hline
\textbf{WNVG - QG} 			& \cellcolor{e5e5e5}0.84 & \cellcolor{e5e5e5}0.94 & \cellcolor{e5e5e5}0.67 \\
\hline
\textbf{WHVG - QG} 			& \cellcolor{cccccc}0.90 & \cellcolor{cccccc}0.96 & \cellcolor{cccccc}\textbf{0.73} \\
\hline
\textbf{\textit{NetF}}		& \cellcolor{b2b2b2}\textbf{0.92} & \cellcolor{b2b2b2}\textbf{0.97} & \cellcolor{b2b2b2}0.68 \\
\hline
\end{tabular}
\label{table:3}
\end{table}

These  results show that different mapping methods capture different properties from the series, as we analyzed in the Section~\ref{subsec:charact_models}, translating into a better clustering result, as we expected. 
If we analyse  only  feature vectors corresponding to one network kind, the first three rows of Table~\ref{table:3}, we note that the WHVGs are the ones that best capture the characteristics of time series models, having high evaluation values, namely, 0.83 for ARI, 0.94 for NMI and 0.63 for AS. 
The last three lines of Table~\ref{table:3} show better results than those obtained using only WHVG features, thus showing that the resulting features of the QGs add information about certain properties of the time series models. 

We still study how these seven sets of  features perform in determining the number of clusters $k=11$  using the ARI, NMI and AS evaluation metrics. 
The results for $k$ obtained from features corresponding to only one kind of network, range from 8 to 13 for ARI, from 11 to 14 for NMI and 3 to 9 for AS. However, when \textit{NetF} is used, we obtain $k=11$ for ARI and NMI and $k=10$ for AS.

\newpage
\section{Classical Features}
\label{app:features}
The above described procedure is applied to two further  sets of features previously proposed in the literature. One is a   set of time series statistical features that has been used in a variety of tasks such as clustering~\cite{Wang2006}, forecasting~\cite{kang2017visualising,talagala2018meta} and generation of time series data~\cite{kang2020gratis}. It comprises sixteen measures calculated using the \texttt{tsfeatures} package~\cite{tsfeatures} of the R CRAN~\cite{R}, namely, frequency and number and length of seasonal periods, strength of trend, "spikiness" of a time series, linearity and curvature, spectral entropy, and  measures based on autocorrelation coefficients of the original series, first-differenced series and second-differenced series. These will be denoted by \textit{tsfeatures} in the remainder of this work. 
The second is denominated canonical feature set, \textit{catch22}~\cite{lubba2019catch22}, has been recently proposed  based on a features library from an interdisciplinary time series analysis literature~\cite{fulcher2017hctsa} and has been used in time series classification tasks~\cite{lubba2019catch22}. There are twenty two measures calculated using the \texttt{Rcatch22}\footnote{\url{https://github.com/hendersontrent/Rcatch22}} package~\cite{catch22} of the R CRAN~\cite{R}, that include properties of the distributions and simple temporal statistics of values in the time series, linear and non-linear autocorrelation, successive differences, scaling of fluctuations, and others.

\newpage
\section{Clustering Time Series with \textit{NetF}}
\label{app:clust_results}

\begin{figure}[hbt!]
	\centering 
	\includegraphics[scale=.4]{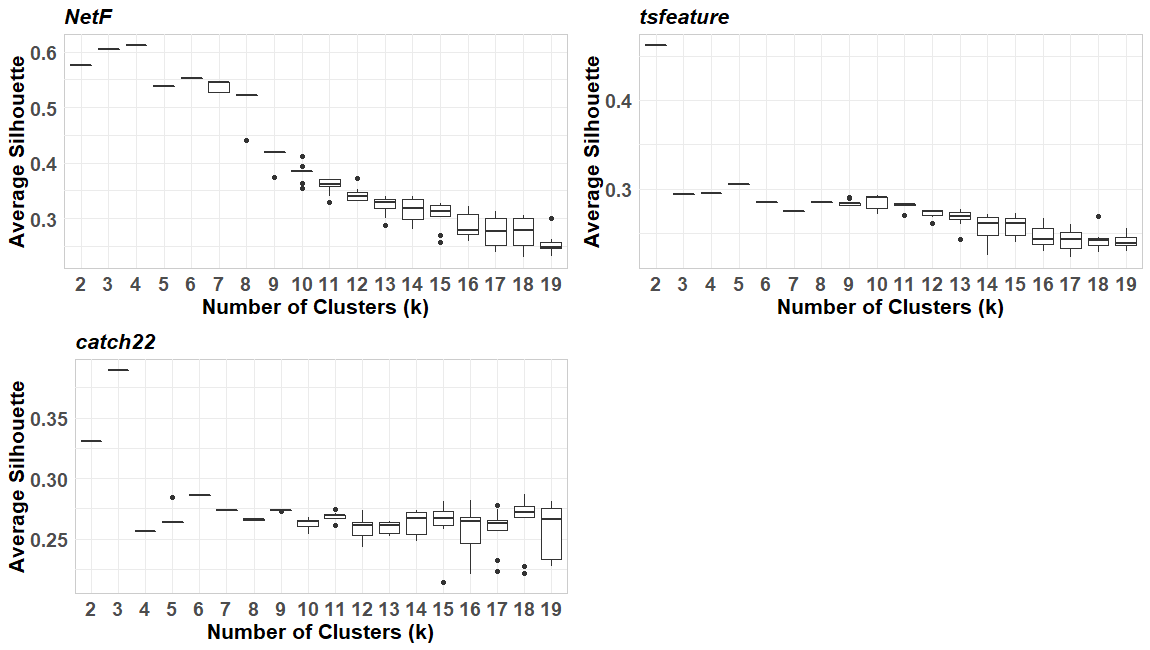}
	\caption{Number of clusters, $k$, for the Production in Brazil data set using the silhouette method for 10 repetitions of the clustering analysis using the 3  features vectors: \textit{NetF}, \textit{catch22} and \textit{tsfeature}.}
	\label{app:Figure_ASbestk}
\end{figure}

%\begin{sidewaystable}[htbp]
\begin{sidewaystable}
\begin{center}
\begin{minipage}{\textheight}

\caption{Clustering evaluation metrics obtained for the two conventional approaches (\textit{tsf.} and \textit{cat.}) and for the proposed approach (\textit{NetF}). 
The number of clusters, $k$, is determined automatically by the evaluation metrics. }

\begin{tabular}{ |l|c|c|c|c|c|c|c|c|c|c|c|c|c| }
\hline
& \multicolumn{1}{|c|}{\textbf{\scriptsize Num. of}} & \multicolumn{3}{c|}{\textbf{Best $k$}} & \multicolumn{3}{c|}{\textbf{ARI}} & \multicolumn{3}{c|}{\textbf{NMI}} &  \multicolumn{3}{c|}{\textbf{AS}}\\
\multicolumn{1}{|c|}{\textbf{Data set}} & \multicolumn{1}{|c|}{\textbf{\scriptsize Classes}} & \multicolumn{3}{c|}{{}} & \multicolumn{3}{c|}{{\scriptsize $[-1,1]$}} & \multicolumn{3}{c|}{{\scriptsize $[0,1]$}} & \multicolumn{3}{c|}{{\scriptsize $[-1,1]$}} \\ 

\cline{3-14}
& \multicolumn{1}{|c|}{\textbf{}} & \multicolumn{1}{c|}{{\scriptsize \textit{tsf.}}} & \multicolumn{1}{c|}{{\scriptsize \textit{cat.}}} & \multicolumn{1}{c|}{{\scriptsize \textit{NetF}}} & \multicolumn{1}{c|}{{\scriptsize \textit{tsf.}}} & \multicolumn{1}{c|}{{\scriptsize \textit{cat.}}} & \multicolumn{1}{c|}{{\scriptsize \textit{NetF}}} & \multicolumn{1}{c|}{{\scriptsize \textit{tsf.}}} & \multicolumn{1}{c|}{{\scriptsize \textit{cat.}}} & \multicolumn{1}{c|}{{\scriptsize \textit{NetF}}} & \multicolumn{1}{c|}{{\scriptsize \textit{tsf.}}} & \multicolumn{1}{c|}{{\scriptsize \textit{cat.}}} & \multicolumn{1}{c|}{{\scriptsize \textit{NetF}}} \\

\hline
{\scriptsize \textbf{18Pairs}} 				  & 18 & 21 & 19 & 21 & 
\textbf{0.51} & 0.41 & 0.39 & 
\textbf{0.90} & 0.87 & 0.87 & 
\textbf{0.40} & 0.37 & 0.28 \\
\hline
{\scriptsize \textbf{M3 data}} 				  & 6 & 9 & 9 & 5 & 
0.14 & 0.14 & 0.14 & 
\textbf{0.22} & 0.21 & 0.18 & 
\textbf{0.31} & 0.26 & 0.26 \\
\hline
{\scriptsize \textbf{CinC\_ECG\_torso}} 	  & 4 & 4 & 4 & 5 & 
0.31 & 0.32 & \textbf{0.45} & 
0.37 & 0.35 & \textbf{0.53} & 
0.23 & 0.19 & \textbf{0.31} \\
\hline
{\scriptsize \textbf{Cricket\_X}} 			  & 12 & 9 & 17 & 10 & 
0.17 & \textbf{0.16} & \textbf{0.16} & 
\textbf{0.31} & 0.30 & 0.29 & 
\textbf{0.20} & 0.15 & 0.10 \\
\hline
{\scriptsize \textbf{ECG5000}} 				  & 5 & 4 & 2 & 2 & 
0.35 & \textbf{0.45} & \textbf{0.45} & 
\textbf{0.37} & 0.35 & 0.32 & 
0.25 & \textbf{0.33} & 0.19 \\
\hline
{\scriptsize \textbf{ElectricDevices}} 		  & 7 & 9 & 3 & 5 & 
0.21 & 0.26 & \textbf{0.28} &
0.30 & \textbf{0.35} & 0.32 & 
\textbf{0.31} & 0.28 & 0.30 \\
\hline
{\scriptsize \textbf{FaceAll}} 				  & 14 & 23 & 14 & 11 &
0.16 & \textbf{0.20} & 0.15 & 
\textbf{0.37} & 0.36 & 0.27 & 
\textbf{0.20} & 0.15 & 0.09 \\
\hline
{\scriptsize \textbf{FordA}} 				  & 2 & 4 & 6 & 3 & 
\textbf{0.21} & 0.12 & 0.13 & 
\textbf{0.30} & 0.21 & 0.13 & 
\textbf{0.31} & 0.14 & 0.21 \\
\hline
{\scriptsize \textbf{InsectWingbeatSound}} 	  & 11 & 6 & 11 & 3 & 
0.09 & \textbf{0.21} & 0.20 & 
0.19 & 0.36 & \textbf{0.39} &
\textbf{0.21} & 0.18 & 0.17 \\
\hline
{\scriptsize \textbf{UWaveGestureLibraryAll}}           & 8 & 16 & 10 & 6 & 
0.21 & 0.21 & 0.21 &
\textbf{0.36} & 0.30 & 0.30 & 
0.19 & \textbf{0.20} & 0.15  \\
\hline
\hline
{\scriptsize \textbf{Synthetic (DGP)}}	            & 11 & 15 & 8 & 11 & 
0.79 & 0.47 & \textbf{0.92} & 
0.91 & 0.69 & \textbf{0.97} & 
0.64 & 0.38 & \textbf{0.68} \\
\hline
\end{tabular}
\label{app:table_bestK}
\end{minipage}
\end{center}
\end{sidewaystable}

\newpage
\section{Clustering Results: UEA \& UCR Time Series Datasets}
\label{app:allclustres}

For some sets of benchmark empirical time series, some features of \textit{tsfeatures} and \textit{catch22} approaches (see Appendix~\ref{app:features}) return missing values, and some have time series with missing values. 
We decided not to consider these sets in our clustering analysis as they are just a few. 
So Tables~\ref{app:table_utsc_res1} to~\ref{app:table_utsc_res3} present the results for 119 sets, out of a total of 129.

\begin{table}[hbt!]
\centering
\caption{Brief description of the empirical time series datasets from UEA \& UCR time series repository~\cite{UCRArchive} and the clustering evaluation metrics obtained for the two conventional approaches (\textit{tsfeatures} and \textit{catch22}) and for the proposed approach (\textit{NetF}). The values reflect the mean of 10 repetitions of the clustering analysis for the ground truth, $k$. 
The values in bold represent the best results of the respective evaluation metric comparing the two approaches.
$M$ represents the size of data set, $T$ the time series length and $k$ the number of classes.}
\begin{tabular}{ |>{\scriptsize}l|>{\scriptsize}r|>{\scriptsize}r|>{\scriptsize}r|>{\scriptsize}c|>{\scriptsize}c|>{\scriptsize}c|>{\scriptsize}c|>{\scriptsize}c|>{\scriptsize}c|>{\scriptsize}c|>{\scriptsize}c|>{\scriptsize}c| }  
\hline
 & & & & \multicolumn{3}{c|}{\textbf{ARI}} & \multicolumn{3}{c|}{\textbf{NMI}} &  \multicolumn{3}{c|}{\textbf{AS}}\\
\multicolumn{1}{|c|}{\textbf{Data set}} & \multicolumn{1}{c|}{\textbf{$M$}} & \multicolumn{1}{c|}{\textbf{$T$}} & \multicolumn{1}{c|}{\textbf{$k$}} & \multicolumn{3}{c|}{{\scriptsize $[-1,1]$}} & \multicolumn{3}{c|}{{\scriptsize $[0,1]$}} & \multicolumn{3}{c|}{{\scriptsize $[-1,1]$}} \\
\cline{5-13}
 & & & & \multicolumn{1}{c|}{{\scriptsize \textit{tsf.}}} & \multicolumn{1}{c|}{{\scriptsize \textit{cat.}}} & \multicolumn{1}{c|}{{\scriptsize \textit{Net.}}} & \multicolumn{1}{c|}{{\scriptsize \textit{tsf.}}} & \multicolumn{1}{c|}{{\scriptsize \textit{cat.}}} & \multicolumn{1}{c|}{{\scriptsize \textit{Net.}}} & \multicolumn{1}{c|}{{\scriptsize \textit{tsf.}}} & \multicolumn{1}{c|}{{\scriptsize \textit{cat.}}} & \multicolumn{1}{c|}{{\scriptsize \textit{Net.}}} \\
 
\hline

{\scriptsize \textbf{ACSF1}}                                & 200 & 1460 & 10 &     0.22 & \textbf{0.32} & 0.17 & 0.50 & \textbf{0.56} & 0.40 & \textbf{0.48} & 0.30 & 0.23\\
\hline
{\scriptsize \textbf{Adiac}}                                & 781 & 176 & 37 &     \textbf{0.21} & \textbf{0.21} & 0.11 & \textbf{0.55} & \textbf{0.55} & 0.43 & 0.20 & \textbf{0.26} & 0.13\\
\hline
{\scriptsize \textbf{ArrowHead}}                            & 211 & 251 & 3 &       0.24 & 0.17 & \textbf{0.34 }& 0.27 & 0.21 & \textbf{0.31} & 0.24 &\textbf{ 0.27} & 0.15\\
\hline
{\scriptsize \textbf{BME}}                                  & 180 & 128 & 3 &       \textbf{0.50 }& 0.36 & 0.40 & \textbf{0.58} & 0.36 & 0.45 & \textbf{0.55} & 0.20 & 0.23\\
\hline
{\scriptsize \textbf{Beef}}                                 & 60 & 470 & 5 &        \textbf{0.09} & 0.05 & 0.04 & \textbf{0.23} & 0.24 & 0.18 & 0.33 & \textbf{0.38} & 0.18\\
\hline
{\scriptsize \textbf{BeetleFly}}                            & 40 & 512 & 2 &        0.55 & 0.10 & \textbf{0.63} & 0.48 & 0.11 & \textbf{0.55} & \textbf{0.34} & 0.26 & 0.16\\
\hline
{\scriptsize \textbf{BirdChicken}}                          & 40 & 512 & 2 &        0.07 & 0.55 & \textbf{0.63} & 0.10 & \textbf{0.56} & 0.53 & \textbf{0.35} & 0.19 & 0.29\\
\hline
{\scriptsize \textbf{CBF}}                                  & 930 & 128 & 3 &       0.31 & \textbf{0.37} & \textbf{0.37} & 0.34 & 0.39 & \textbf{0.40 }& \textbf{0.22} & 0.19 & 0.13\\
\hline
{\scriptsize \textbf{Car}}                                  & 120 & 577 & 4 &       \textbf{0.25} & 0.16 & 0.20 & \textbf{0.35} & 0.23 & 0.29 & 0.22 &\textbf{ 0.28} & 0.18\\
\hline
{\scriptsize \textbf{Chinatown}}                            & 365 & 24 & 2 &        \textbf{0.33} & 0.29 & -0.05 & \textbf{0.29} & 0.21 & 0.03 & 0.23 & \textbf{0.34} & 0.27\\
\hline
{\scriptsize \textbf{ChlorineConcentration}}                & 4307 & 166 & 3 &      0.00 & 0.00 & 0.00 & 0.00 & 0.00 & 0.00 & \textbf{0.46} & 0.42 & 0.17\\
\hline
{\scriptsize \textbf{CinCECGTorso}}                         & 1420 & 1639 & 4 &     0.31 & 0.32 & \textbf{0.45} & 0.37 & 0.35 & \textbf{0.52} & 0.23 & 0.19 & \textbf{0.31}\\
\hline
{\scriptsize \textbf{Coffee}}                               & 56 & 286 & 2 &        0.61 & \textbf{1.00} & 0.45 & 0.54 & \textbf{1.00} & 0.42 & \textbf{0.32} & 0.21 & 0.13\\
\hline
{\scriptsize \textbf{Computers}}                            & 500 & 720 & 2 &       0.00 & 0.00 & \textbf{0.07} & 0.00 & 0.00 & \textbf{0.05} & \textbf{0.46} & 0.17 & 0.27\\
\hline
{\scriptsize \textbf{CricketX}}                             & 780 & 300 & 12 &      0.15 & 0.15 & \textbf{0.16} & \textbf{0.32} & 0.28 & 0.30 & \textbf{0.20} & 0.16 & 0.10\\
\hline
{\scriptsize \textbf{CricketY}}                             & 780 & 300 & 12 &      0.13 & \textbf{0.14} & 0.08 & 0.27 & \textbf{0.28} & 0.21 & \textbf{0.19} & 0.17 & 0.09\\
\hline
{\scriptsize \textbf{CricketZ}}                             & 780 & 300 & 12 &      0.15 & \textbf{0.16} & 0.14 & \textbf{0.33} & 0.28 & 0.27 & \textbf{0.19} & 0.16 & 0.10\\
\hline
{\scriptsize \textbf{Crop}}                                 & 24000 & 46 & 24 &     \textbf{0.19} & 0.16 & 0.09 & \textbf{0.39 }& 0.34 & 0.26 & \textbf{0.19 }& 0.18 & 0.09\\
\hline
{\scriptsize \textbf{DiatomSizeReduction}}                  & 322 & 345 & 4 &       0.67 & 0.11 & \textbf{0.69} & \textbf{0.67} & 0.26 & \textbf{0.67} & \textbf{0.34} & \textbf{0.34} & 0.26\\
\hline
{\scriptsize \textbf{DistalPhalanxOutlineAgeG}}             & 539 &  80 & 3 &       \textbf{0.45} & 0.44 & 0.20 & \textbf{0.35} & 0.34 & \textbf{0.35} & \textbf{0.47} & \textbf{0.47} & 0.44\\
\hline
{\scriptsize \textbf{DistalPhalanxOutlineCorr}}             & 876 & 80 & 2 &        0.00 & 0.00 & \textbf{0.01} & 0.00 & 0.00 & \textbf{0.01} & 0.45 & 0.42 & \textbf{0.69}\\
\hline
{\scriptsize \textbf{DistalPhalanxTW}}                      & 539 & 80 & 6 &        0.41 & \textbf{0.45} & 0.28 & \textbf{0.47} & 0.45 & 0.43 & 0.31 & \textbf{0.42} & 0.11\\
\hline
{\scriptsize \textbf{ECG200}}                               & 200 & 96 & 2 &        \textbf{0.25} & 0.07 & 0.03 & \textbf{0.16} & 0.05 & 0.04 & \textbf{0.30} & 0.19 & 0.16\\
\hline
{\scriptsize \textbf{ECG5000}}                              & 5000 & 140 & 5 &      0.29 & 0.28 & \textbf{0.31} & \textbf{0.32} & 0.29 & 0.30 & \textbf{0.24} & \textbf{0.24} & 0.16\\
\hline
{\scriptsize \textbf{ECGFiveDays}}                          & 884 & 136 & 2 &       \textbf{0.02} & 0.00 & 0.00 & \textbf{0.02} & 0.00 & 0.01 & \textbf{0.36} & 0.30 & 0.25\\
\hline
{\scriptsize \textbf{ElectricDevices}}                      & 16575 & 96 & 7 &      0.20 & \textbf{0.21} & 0.19 & \textbf{0.30} & 0.29 & 0.29 & \textbf{0.33} & 0.25 & 0.27 \\
\hline
{\scriptsize \textbf{EOGHorizontalSignal}}                  & 724 & 1250 & 12 &         \textbf{0.18} & 0.16 & 0.12 & \textbf{0.38} & 0.33 & 0.27 & \textbf{0.27} & 0.22 & 0.13\\
\hline
{\scriptsize \textbf{EOGVerticalSignal}}                    & 724 & 1250 & 12 &         0.10 & \textbf{0.13} & 0.06 & 0.26 & \textbf{0.28} & 0.17 & \textbf{0.21} & 0.20 & 0.11\\
\hline
{\scriptsize \textbf{Earthquakes}}                          & 461 & 512 & 2 &        \textbf{0.00} & -0.03 & -0.07 & 0.00 & \textbf{0.07} & 0.04 & 0.27 & 0.18 & \textbf{0.52}\\
\hline
{\scriptsize \textbf{EthanolLevel}}                         & 1004 & 1751 & 4 &         \textbf{0.01} & 0.00 & \textbf{0.01} & 0.01 & 0.00 & \textbf{0.02} & 0.15 & \textbf{0.27} & 0.15\\
\hline
{\scriptsize \textbf{FaceAll}}                              & 2250 & 131 & 14 &         0.15 & \textbf{0.21} & 0.15 & 0.33 & \textbf{0.36} & 0.29 & \textbf{0.22} & 0.15 & 0.09\\
\hline
{\scriptsize \textbf{FaceFour}}                             & 112 & 350 & 4 &       0.26 & 0.36 & \textbf{0.46} & 0.32 & 0.45 & \textbf{0.54} & \textbf{0.39} & 0.22 & 0.23\\
\hline
{\scriptsize \textbf{FacesUCR}}                             & 2250 & 131 & 14 &      0.14 & \textbf{0.19} & 0.15 & 0.33 & \textbf{0.37} & 0.28 & \textbf{0.22} & 0.15 & 0.09\\
 \hline
 {\scriptsize \textbf{FiftyWords}}                           & 905 & 270 & 50 &      \textbf{0.19} & \textbf{0.19} & 0.09 & \textbf{0.57} & 0.54 & 0.44 & \textbf{0.18} & 0.17 & 0.11\\
\hline
{\scriptsize \textbf{Fish}}                                 & 350 & 463 & 7 &       \textbf{0.19} & 0.12 & 0.17 & 0.29 & 0.21 & \textbf{0.30} & 0.20 & \textbf{0.25} & 0.13\\

\hline
\end{tabular}
\label{app:table_utsc_res1}
\end{table}

\begin{table}[hbt!]
\centering
\caption{(\textit{cont.}) Brief description of the empirical time series datasets from UEA \& UCR time series repository~\cite{UCRArchive} and the clustering evaluation metrics obtained for the two conventional approaches (\textit{tsfeatures} and \textit{catch22}) and for the proposed approach (\textit{NetF}). The values reflect the mean of 10 repetitions of the clustering analysis for the ground truth, $k$. 
The values in bold represent the best results of the respective evaluation metric comparing the two approaches.
$M$ represents the size of data set, $T$ the time series length and $k$ the number of classes.}
\begin{tabular}{ |>{\scriptsize}l|>{\scriptsize}r|>{\scriptsize}r|>{\scriptsize}r|>{\scriptsize}c|>{\scriptsize}c|>{\scriptsize}c|>{\scriptsize}c|>{\scriptsize}c|>{\scriptsize}c|>{\scriptsize}c|>{\scriptsize}c|>{\scriptsize}c| }  

\hline
 & & & & \multicolumn{3}{c|}{\textbf{ARI}} & \multicolumn{3}{c|}{\textbf{NMI}} &  \multicolumn{3}{c|}{\textbf{AS}}\\
\multicolumn{1}{|c|}{\textbf{Data set}} & \multicolumn{1}{c|}{\textbf{$M$}} & \multicolumn{1}{c|}{\textbf{$T$}} & \multicolumn{1}{c|}{\textbf{$k$}} & \multicolumn{3}{c|}{{\scriptsize $[-1,1]$}} & \multicolumn{3}{c|}{{\scriptsize $[0,1]$}} & \multicolumn{3}{c|}{{\scriptsize $[-1,1]$}} \\
\cline{5-13}
 & & & & \multicolumn{1}{c|}{{\scriptsize \textit{tsf.}}} & \multicolumn{1}{c|}{{\scriptsize \textit{cat.}}} & \multicolumn{1}{c|}{{\scriptsize \textit{Net.}}} & \multicolumn{1}{c|}{{\scriptsize \textit{tsf.}}} & \multicolumn{1}{c|}{{\scriptsize \textit{cat.}}} & \multicolumn{1}{c|}{{\scriptsize \textit{Net.}}} & \multicolumn{1}{c|}{{\scriptsize \textit{tsf.}}} & \multicolumn{1}{c|}{{\scriptsize \textit{cat.}}} & \multicolumn{1}{c|}{{\scriptsize \textit{Net.}}} \\
\hline

{\scriptsize \textbf{FordA}}                                & 4921 & 500 & 2 &       \textbf{0.19} & 0.01 & 0.01 & \textbf{0.27} & 0.01 & 0.01 & \textbf{0.53} & 0.33 & 0.29\\
\hline
{\scriptsize \textbf{FordB}}                                & 4446 & 500 & 2 &      \textbf{0.27} & 0.07 & 0.02 & \textbf{0.31} & 0.07 & 0.02 & \textbf{0.48} & 0.29 & 0.22\\
\hline
{\scriptsize \textbf{FreezerRegularTrain}}                  & 3000 & 301 & 2 &      0.22 & 0.27 & \textbf{0.31} & 0.19 & 0.21 & \textbf{0.24} & \textbf{0.52} & 0.50 & 0.11\\
\hline
{\scriptsize \textbf{FreezerSmallTrain}}                    & 2878 & 301 & 2 &      0.22 & 0.27 & \textbf{0.32} & 0.19 & 0.21 & \textbf{0.24} & \textbf{0.52} & 0.50 & 0.11\\
\hline
{\scriptsize \textbf{Fungi}}                                & 204 & 201 & 18 &      \textbf{0.77} & 0.72 & 0.40 & \textbf{0.92} & 0.90 & 0.68 & \textbf{0.48} & 0.43 & 0.13\\
\hline
{\scriptsize \textbf{GestureMidAirD1}}                      & 338 & 360 & 26 &               0.27 & 0.25 & \textbf{0.31} & 0.62 & 0.59 & \textbf{0.63} & \textbf{0.23} & \textbf{0.23} & 0.15\\
\hline
{\scriptsize \textbf{GestureMidAirD2}}                      & 338 & 360 & 26&               0.24 & 0.25 & \textbf{0.26} & 0.60 & 0.60 & \textbf{0.61} & 0.21 & \textbf{0.23} & 0.15\\
\hline
{\scriptsize \textbf{GestureMidAirD3}}                      & 338 & 360 & 26 &              \textbf{0.18} & 0.14 & 0.16 & \textbf{0.52} & 0.48 & 0.50 & \textbf{0.19} & \textbf{0.19} & 0.13\\
\hline
{\scriptsize \textbf{GesturePebbleZ1}}                      & 304 & $[100, 455]$ & 6 &      0.15 & 0.16 & \textbf{0.23} & 0.23 & 0.22 & \textbf{0.35} & 0.16 & 0.15 & \textbf{0.18}\\
\hline
{\scriptsize \textbf{GesturePebbleZ2}}                      & 304 & $[100, 455]$ & 6 &          0.15 & 0.16 & \textbf{0.23 }& 0.23 & 0.22 & \textbf{0.35} & 0.16 & 0.15 & \textbf{0.18}\\
\hline
{\scriptsize \textbf{GunPoint}}                             & 200 & 150 & 2 &           0.00 & 0.00 & \textbf{0.19} & 0.06 & 0.00 & \textbf{0.30} & \textbf{0.70} & 0.28 & 0.27\\
\hline
{\scriptsize \textbf{GunPointAgeSpan}}                      & 451 & 150 & 2 &       0.01 & \textbf{0.10} & 0.00 & 0.02 & \textbf{0.07} & 0.00 & \textbf{0.48} & 0.23 & 0.30\\
\hline
{\scriptsize \textbf{GunPointMaleVersusFemale}}             & 451 & 150 & 2 &       0.07 & 0.05 & \textbf{0.32} & 0.13 & 0.04 & \textbf{0.37} & \textbf{0.48} & 0.23 & 0.30\\
\hline
{\scriptsize \textbf{GunPointOldVersusYoung}}               & 451 & 150 & 2 &       0.00 & 0.08 & \textbf{0.20} & 0.00 & 0.06 & \textbf{0.27} & \textbf{0.48} & 0.23 & 0.30\\
\hline
{\scriptsize \textbf{Ham}}                                  & 214 & 431 & 2 &       0.00 & 0.00 & \textbf{0.01} & 0.00 & \textbf{0.01} & \textbf{0.01} & \textbf{0.33} & 0.16 & 0.18\\
\hline
{\scriptsize \textbf{HandOutlines}}                         & 1370 & 2709 & 2 &         0.05 & 0.02 & \textbf{0.06} & 0.02 & 0.00 & \textbf{0.11} & 0.37 & 0.50 & \textbf{0.53}\\
\hline
{\scriptsize \textbf{Haptics}}                              & 463 & 1092 & 5 &      0.04 & 0.03 & \textbf{0.07} & 0.07 & 0.07 & \textbf{0.10} & \textbf{0.38} & 0.28 & 0.15\\
\hline
{\scriptsize \textbf{Herring}}                              &  128 & 512 & 2 &      \textbf{0.00} & -0.01 & -0.01 & 0.00 & 0.00 & 0.00 & \textbf{0.22} & 0.16 & 0.17\\
\hline
{\scriptsize \textbf{HouseTwenty}}                          & 135 & 3000 & 2 &      -0.01 & \textbf{0.64} & 0.10 & 0.01 & \textbf{0.61} & 0.09 & \textbf{0.40} & 0.21 & 0.29\\
\hline
{\scriptsize \textbf{InlineSkate}}                          & 650 & 1882 & 7 &      \textbf{0.10} & 0.02 & 0.08 & \textbf{0.22} & 0.07 & 0.17 & \textbf{0.29} & 0.18 & 0.16\\
\hline
{\scriptsize \textbf{InsectEPGRegularTrain}}                & 311 & 601 & 3 &       0.50 & 0.43 & \textbf{0.55} & \textbf{0.65} & 0.44 & 0.61 & \textbf{0.37} & 0.22 & 0.19\\
\hline
{\scriptsize \textbf{InsectEPGSmallTrain}}                  & 266 & 601 & 3 &       0.50 & 0.44 & \textbf{0.52} & \textbf{0.65} & 0.46 & 0.61 & \textbf{0.37} & 0.21 & 0.19\\
\hline
{\scriptsize \textbf{InsectWingbeatSound}}                  & 2200 & 256 & 11 &      0.07 & \textbf{0.21} & 0.17 & 0.18 & \textbf{0.37} & 0.32 & \textbf{0.19} & 0.18 & 0.11\\
\hline
{\scriptsize \textbf{ItalyPowerDemand}}                     & 1096 & 24 & 2 &       \textbf{0.04} & 0.01 & 0.03 & \textbf{0.05} & 0.01 & 0.03 & 0.38 & \textbf{0.40} & 0.27\\
\hline
{\scriptsize \textbf{LargeKitchenAppliances}}               & 750 & 720 & 3 &          \textbf{0.21} &  0.06 & 0.00 & \textbf{0.23} & 0.05 & 0.01 & \textbf{ 0.35} & 0.23 & 0.30\\
\hline
{\scriptsize \textbf{Lightning2}}                           & 121 & 637 & 2 &        \textbf{0.07} &0.02 & 0.05 & \textbf{0.14} & 0.04 & 0.07 & \textbf{0.42} & 0.28 & 0.19\\
\hline
{\scriptsize \textbf{Lightning7}}                           & 143 & 319 & 7 &        \textbf{0.22} & 0.21 & 0.18 & \textbf{0.39} & \textbf{0.39} & 0.35 & \textbf{0.24} & \textbf{0.24} & 0.14\\
\hline
{\scriptsize \textbf{Mallat}}                               & 2400 & 1024 & 8 &      \textbf{0.70} & 0.69 & 0.53 & 0.80 & \textbf{0.83} & 0.65 & \textbf{0.35} & 0.32 & 0.13\\
\hline
{\scriptsize \textbf{Meat}}                                 & 120 & 448 & 3 &       \textbf{0.45} & \textbf{0.45} & 0.17 & 0.46 & \textbf{0.63} & 0.18 & 0.29 & \textbf{0.43} & 0.13\\
\hline
{\scriptsize \textbf{MedicalImages}}                        & 1141 & 99 & 10 &       \textbf{0.10} & 0.03 & 0.06 & \textbf{0.28} & 0.19 & 0.17 & \textbf{0.31} & 0.21 & 0.17\\
\hline
{\scriptsize \textbf{MiddlePhalanxOutlineAgeG}}         & 554 & 80 & 3 &        0.42 & \textbf{0.43} & 0.42 & 0.39 & 0.39 & 0.39 & \textbf{0.56} & 0.49 & 0.40\\
\hline
{\scriptsize \textbf{MiddlePhalanxOutlineCorr}}          & 891 & 80 & 2 &       -0.01 & \textbf{0.00} & -0.01 & \textbf{0.01} & 0.00 & \textbf{0.01 }& 0.47 & 0.41 & \textbf{0.71}\\
\hline
{\scriptsize \textbf{MiddlePhalanxTW}}                      & 553 & 80 & 6 &        0.34 & \textbf{0.57} & 0.24 & 0.40 & \textbf{0.43} & 0.39 & 0.28 & \textbf{0.47} & 0.14\\
\hline
{\scriptsize \textbf{MixedShapesRegularTrain}}              & 2925 & 1024 & 5 &      0.44 & 0.23 & \textbf{0.52} & 0.49 & 0.26 & \textbf{0.55} & \textbf{0.31} & 0.20 & 0.21\\
\hline
{\scriptsize \textbf{MixedShapesSmallTrain}}                & 2525 & 1024 & 5 &      0.45 & 0.23 & \textbf{0.51} & 0.49 & 0.25 & \textbf{0.54} & \textbf{0.31} & 0.20 & 0.21\\
\hline
{\scriptsize \textbf{MoteStrain}}                           & 1272 & 84 & 2 &       0.01 & 0.02 & \textbf{0.17} & 0.01 & 0.01 & \textbf{0.17} & \textbf{0.48} & 0.25 & 0.20\\
\hline
{\scriptsize \textbf{NonInvasiveFetalECGThor1}}           & 3765 & 750 & 42 &       \textbf{0.51} & 0.27 & 0.07 & \textbf{0.76} & 0.58 & 0.30 & \textbf{0.22} & 0.20 & 0.09\\
\hline
{\scriptsize \textbf{NonInvasiveFetalECGThor2}}           & 3765 & 750 & 42 &       \textbf{0.54} & 0.36 & 0.11 & \textbf{0.79} & 0.67 & 0.36 & \textbf{0.25} & 0.23 & 0.11\\
\hline
{\scriptsize \textbf{OSULeaf}}                              & 442 & 427 & 6 &       0.29 & 0.28 & \textbf{0.49} & 0.42 & 0.38 & \textbf{0.54} & \textbf{0.17} & 0.15 & 0.16\\
\hline
{\scriptsize \textbf{OliveOil}}                             & 60 & 570 & 4&         \textbf{0.29} & 0.18 & 0.10 & \textbf{0.36} & 0.27 & 0.17 & 0.30 & \textbf{0.32} & 0.10\\
\hline
{\scriptsize \textbf{PLAID}}                                & 1074 & $[100, 1344]$ & 11 &        \textbf{0.38} & 0.27 & 0.25 & \textbf{0.51} & 0.41 & 0.38 & \textbf{0.31} & 0.22 & 0.18\\
\hline
{\scriptsize \textbf{PhalangesOutlinesCorrect}}             & 2658 & 80 & 2 &               \textbf{0.01} & \textbf{0.01} & 0.00 & 0.00 & 0.00 & 0.00 & 0.33 & 0.37 & \textbf{0.70}\\
\hline
{\scriptsize \textbf{Phoneme}}                              & 2110 & 1024 & 39 &            0.06 & 0.07 & \textbf{0.09} & 0.29 & 0.31 & \textbf{0.34} & 0.14 & \textbf{0.13} & \textbf{0.13}\\
\hline
{\scriptsize \textbf{PickupGestureWiimoteZ}}                & 100 & $[29, 361]$ & 10 &             0.22 & 0.24 & \textbf{0.29} & 0.50 & 0.50 & \textbf{0.55} & 0.18 & 0.18 & \textbf{0.19}\\
\hline
{\scriptsize \textbf{PigAirwayPressure}}                    & 312 & 2000 & 52 &                 0.11 & 0.04 & \textbf{0.13} & 0.63 & 0.57 & \textbf{0.64} & \textbf{0.26} & 0.20 & 0.17\\

\hline
\end{tabular}
\label{app:table_utsc_res2}
\end{table}

\begin{table}[hbt!]
\centering
\caption{(\textit{cont.}) Brief description of the empirical time series datasets from UEA \& UCR time series repository~\cite{UCRArchive} and the clustering evaluation metrics obtained for the two conventional approaches (\textit{tsfeatures} and \textit{catch22}) and for the proposed approach (\textit{NetF}). The values reflect the mean of 10 repetitions of the clustering analysis for the ground truth, $k$. 
The values in bold represent the best results of the respective evaluation metric comparing the two approaches.
$M$ represents the size of data set, $T$ the time series length and $k$ the number of classes.}
\begin{tabular}{ |>{\scriptsize}l|>{\scriptsize}r|>{\scriptsize}r|>{\scriptsize}r|>{\scriptsize}c|>{\scriptsize}c|>{\scriptsize}c|>{\scriptsize}c|>{\scriptsize}c|>{\scriptsize}c|>{\scriptsize}c|>{\scriptsize}c|>{\scriptsize}c| }  
\hline
 & & & & \multicolumn{3}{c|}{\textbf{ARI}} & \multicolumn{3}{c|}{\textbf{NMI}} &  \multicolumn{3}{c|}{\textbf{AS}}\\
\multicolumn{1}{|c|}{\textbf{Data set}} & \multicolumn{1}{c|}{\textbf{$M$}} & \multicolumn{1}{c|}{\textbf{$T$}} & \multicolumn{1}{c|}{\textbf{$k$}} & \multicolumn{3}{c|}{{\scriptsize $[-1,1]$}} & \multicolumn{3}{c|}{{\scriptsize $[0,1]$}} & \multicolumn{3}{c|}{{\scriptsize $[-1,1]$}} \\
\cline{5-13}
 & & & & \multicolumn{1}{c|}{{\scriptsize \textit{tsf.}}} & \multicolumn{1}{c|}{{\scriptsize \textit{cat.}}} & \multicolumn{1}{c|}{{\scriptsize \textit{Net.}}} & \multicolumn{1}{c|}{{\scriptsize \textit{tsf.}}} & \multicolumn{1}{c|}{{\scriptsize \textit{cat.}}} & \multicolumn{1}{c|}{{\scriptsize \textit{Net.}}} & \multicolumn{1}{c|}{{\scriptsize \textit{tsf.}}} & \multicolumn{1}{c|}{{\scriptsize \textit{cat.}}} & \multicolumn{1}{c|}{{\scriptsize \textit{Net.}}} \\
 
 \hline
 
{\scriptsize \textbf{PigArtPressure}}                       & 312 & 2000 & 52 &         0.40 & \textbf{0.45} & 0.44 & 0.79 & \textbf{0.82} & 0.81 & \textbf{0.24} & \textbf{0.24} & 0.22\\
\hline
{\scriptsize \textbf{PigCVP}}                               & 312 & 2000 & 52 &      0.13 & 0.18 & \textbf{0.20} & 0.64 & \textbf{0.68} & \textbf{0.68} & \textbf{0.20} & 0.14 & 0.16\\
\hline
{\scriptsize \textbf{Plane}}                                & 210 & 144 & 7 &       \textbf{0.98} & 0.86 & 0.97 & \textbf{0.98} & 0.90 & 0.97 & \textbf{0.58} & 0.34 & 0.37\\
\hline
{\scriptsize \textbf{PowerCons}}                            & 360 & 144 & 2 &        0.13 & 0.00 & \textbf{0.25} & 0.11 & 0.00 & \textbf{0.22} & \textbf{0.22} & 0.20 & \textbf{0.22}\\
\hline
{\scriptsize \textbf{ProximalPhalanxOutlineAg}}       & 605 & 80 & 3 & 0.55 &       \textbf{0.57} & 0.35 & 0.53 & \textbf{0.56} & 0.45 & 0.49 & 0.40 & \textbf{0.50}\\
\hline
{\scriptsize \textbf{ProximalPhalanxOutlineCo}}        & 891 & 80 & 2 &             0.05 & \textbf{0.06} & 0.04 & 0.07 & \textbf{0.08} & 0.05 & 0.44 & 0.49 & \textbf{0.73}\\
\hline
{\scriptsize \textbf{ProximalPhalanxTW}}                    & 605 & 80 & 6  &        0.43 & \textbf{0.47} & 0.34 & 0.56 & \textbf{0.58} & 0.48 & 0.29 & \textbf{0.41} & 0.15\\
\hline
{\scriptsize \textbf{RefrigerationDevices}}                 & 750 & 720 & 3 &       0.02 & 0.02 & 0.02 & 0.02 & 0.02 & 0.02 & \textbf{0.41} & 0.27 & 0.27\\
\hline
{\scriptsize \textbf{Rock}}                                 & 70 & 2844 & 4 &       \textbf{0.36} & 0.16 & 0.35 & \textbf{0.51} & 0.27 & 0.49 & \textbf{0.25} & 0.23 & 0.24\\
\hline
{\scriptsize \textbf{ScreenType}}                           & 750 & 720 & 3 &       0.03 & \textbf{0.04} & 0.01 & 0.03 & \textbf{0.04} & 0.01 & \textbf{0.31} & 0.16 & 0.22\\
\hline
{\scriptsize \textbf{SemgHandGenderCh2}}                    & 900 & 1500 & 2  &      0.00 & 0.00 & \textbf{0.01} & \textbf{0.01} & 0.00 & 0.00 & 0.29 & 0.23 & \textbf{0.31}\\
\hline
{\scriptsize \textbf{SemgHandMovementCh2}}                  & 900 & 1500 &  6 &    0.06 & 0.04 & \textbf{0.10} & 0.14 & 0.12 & \textbf{0.20} & \textbf{0.24} & 0.23 & 0.22\\
\hline
{\scriptsize \textbf{SemgHandSubjectCh2}}                   & 900 & 1500 & 5 &      0.16 & \textbf{0.26} & 0.18 & 0.25 & \textbf{0.33} & 0.28 & 0.25 & 0.24 & \textbf{0.26}\\
\hline
{\scriptsize \textbf{ShakeGestureWiimoteZ}}                 & 100 & $[40, 385]$ & 10 &        0.54 & 0.45 & \textbf{0.56 }& 0.72 & 0.67 & \textbf{0.74} & \textbf{0.29} & 0.20 & 0.27\\
\hline
{\scriptsize \textbf{ShapeletSim}}                          & 200 & 500 & 2 &                   0.10 & \textbf{1.00} & 0.85 & 0.08 & \textbf{1.00} & 0.78 & 0.18 & \textbf{0.33} & 0.17\\
\hline
{\scriptsize \textbf{ShapesAll}}                            & 1200 & 512 & 60 &                  \textbf{0.37} & 0.28 & 0.23 & \textbf{0.70} & 0.65 & 0.60 & \textbf{0.26} & 0.22 & 0.13\\
\hline
{\scriptsize \textbf{SmallKitchenAppliances}}               & 750 & 720 & 3 &       \textbf{0.19} & 0.08 & 0.18 & 0.19 & 0.07 & \textbf{0.20} & 0.32 & 0.25 & \textbf{0.50}\\
\hline
{\scriptsize \textbf{SonyAIBORobotSurface1}} 			    & 621 & 70 & 2  &       \textbf{0.58} & 0.42 & 0.56 & \textbf{0.53} & 0.47 & 0.47 & \textbf{0.34} & 0.26 & 0.25\\
\hline
{\scriptsize \textbf{SonyAIBORobotSurface2}} 			    & 980 & 65 & 2 &        \textbf{0.55} & 0.37 & 0.00 & \textbf{0.47} & 0.28 & 0.01 & \textbf{0.37} & 0.22 & 0.14\\
\hline
{\scriptsize \textbf{StarLightCurves}} 			            & 9236 & 1024 & 3 &         0.49 & 0.43 & \textbf{0.65} & \textbf{0.57} & 0.56 & 0.53 & \textbf{0.42} & 0.34 & 0.32\\
\hline
{\scriptsize \textbf{Strawberry}} 			                & 983 & 235 & 2 &        -0.05 & -0.02 & \textbf{0.01} & \textbf{0.11} & 0.09 & 0.03 & \textbf{0.52 }& 0.35 & 0.16\\
\hline
{\scriptsize \textbf{SwedishLeaf}} 			                & 1125 & 128 & 15 &         \textbf{0.46} & 0.31 & 0.41 & \textbf{0.68} & 0.53 & 0.60 & \textbf{0.25} & 0.23 & 0.16\\
\hline
{\scriptsize \textbf{Symbols}} 			                    & 1020 & 398 & 6 &      0.69 & 0.65 & \textbf{0.77} & 0.78 & 0.79 & \textbf{0.84} & 0.40 & \textbf{0.49} & 0.35\\
\hline
{\scriptsize \textbf{SyntheticControl}} 			        & 600 & 60 & 6 &        0.57 & \textbf{0.61} & 0.43 & 0.71 & \textbf{0.74} & 0.50 & \textbf{0.34} & 0.20 & 0.14\\
\hline
{\scriptsize \textbf{ToeSegmentation1}} 			        & 268 & 277 & 2 &       0.01 & 0.00 & \textbf{0.05} & 0.01 & 0.00 & \textbf{0.05} & 0.22 & \textbf{0.23} & 0.20\\
\hline
{\scriptsize \textbf{ToeSegmentation2}} 			        & 166 & 343 & 2 &       0.12 & 0.07 & \textbf{0.37} & 0.06 & 0.03 & \textbf{0.26} & 0.30 & 0.18 & \textbf{0.37}\\
\hline
{\scriptsize \textbf{Trace}} 			                    & 200 & 275 & 4 &       \textbf{1.00} & 0.73 & 0.63 & \textbf{1.00} & 0.79 & 0.70 & \textbf{0.65} & 0.35 & 0.22\\
\hline
{\scriptsize \textbf{TwoLeadECG}} 			                & 1162 & 82 & 2 &       0.00 & 0.01 & \textbf{0.71} & 0.00 & 0.01 & \textbf{0.61 }& \textbf{0.60} & 0.16 & 0.19\\
\hline
{\scriptsize \textbf{TwoPatterns}} 			                & 5000 & 128 & 4 &      \textbf{0.14} & 0.00 & 0.01 & \textbf{0.17} & 0.00 & 0.01 & \textbf{0.16} & 0.15 & 0.11\\
\hline
{\scriptsize \textbf{UMD}} 			                        & 180 & 150 & 3  &      \textbf{0.48} & 0.17 & 0.35 & \textbf{0.53} & 0.20 & 0.38 & \textbf{0.41} & 0.24 & 0.21\\
\hline
{\scriptsize \textbf{UWaveGestureLibraryAll}} 			    & 4478 & 945 & 8 &       0.17 & \textbf{0.20} & 0.18 & 0.27 & \textbf{0.28} & \textbf{0.28} & \textbf{0.20} & 0.19 & 0.12\\
\hline
{\scriptsize \textbf{UWaveGestureLibraryX}} 			    & 4478 & 315 & 8&        0.18 & 0.19 & \textbf{0.23} & 0.30 & 0.29 & \textbf{0.33} & 0.19 & \textbf{0.20} & 0.15\\
\hline
{\scriptsize \textbf{UWaveGestureLibraryY}} 			    & 4478 & 315 & 8&        \textbf{0.22} & 0.16 & 0.14 & \textbf{0.36} & 0.25 & 0.25 & \textbf{0.22} & 0.20 & 0.15\\
\hline
{\scriptsize \textbf{Wafer}} 			                    & 7164 & 152 & 2 &      -0.02 & -0.04 & \textbf{0.99} & 0.00 & 0.02 & \textbf{0.96} & \textbf{0.56} & 0.33 & 0.51\\
\hline
{\scriptsize \textbf{Wine}} 			                    & 111 & 234 & 2 &       \textbf{0.03} & -0.01 & 0.01 & \textbf{0.03} & 0.00 & 0.01 & 0.39 & \textbf{0.43} & 0.22\\
\hline
{\scriptsize \textbf{WordSynonyms}} 			            & 905 & 270 & 25 &      \textbf{0.14} & 0.10 & 0.05 & \textbf{0.41} & 0.34 & 0.26 & \textbf{0.21} & 0.18 & 0.11\\
\hline
{\scriptsize \textbf{Worms}} 			                    & 258 & 900 & 5 &        \textbf{0.19} & 0.14 & 0.13 & \textbf{0.24} & 0.22 & 0.22 & \textbf{0.26} & 0.17 & 0.22\\
\hline
{\scriptsize \textbf{WormsTwoClass}} 			            & 258 & 900 & 2 &            0.04 & 0.07 & \textbf{0.12} & 0.02 & 0.04 & \textbf{0.08 }& \textbf{0.34} & 0.19 & 0.29\\
\hline
{\scriptsize \textbf{Yoga}} 			                    & 3300 & 426 &   2&        0.00 & 0.00 & 0.00 & 0.00 & 0.00 & 0.00 & \textbf{0.29} & 0.20 & 0.27\\

\hline
\hline
{ \textbf{Mean}} 			                      & & &  &            \textbf{0.24} & 0.22 & \textbf{0.24} & \textbf{0.33} & 0.30 & 0.32 & \textbf{0.33} & 0.26 & 0.23 \\
\hline
{ \textbf{Win}} 			                      & & &  &            46 & 28 & \textbf{47} & \textbf{48} & 28 & 43 & \textbf{82} & 26 & 16 \\
\hline
{ \textbf{Win (\%)}} 			            & & &  &            38.66 & 23.53 & \textbf{39.50} & \textbf{40.34} & 23.53 & 36.13 & \textbf{68.91} & 21.85 & 13.45 \\

\hline
\end{tabular}
\label{app:table_utsc_res3}
\end{table}

\end{appendices}

\end{document}